\documentclass[11pt]{article}
\usepackage[a4paper,top=3cm,bottom=3cm,left=2.6cm,right=2.6cm]{geometry}
\usepackage{times}  
\usepackage[numbers]{natbib}
\usepackage{float}
\usepackage[T1]{fontenc}
\usepackage[utf8]{inputenc}
\usepackage{microtype}
\usepackage{textcomp}
\usepackage{amsmath,amssymb}
\usepackage{graphicx}
\usepackage{authblk}      
\usepackage{titlesec}
\usepackage{xcolor}
\usepackage[colorlinks=true, linkcolor=blue, citecolor=blue, urlcolor=blue]{hyperref}
\usepackage{cleveref} 
\usepackage{fancyhdr}
\usepackage{lastpage}
\usepackage{enumitem}
\usepackage{caption}
\usepackage{booktabs}
\usepackage{comment}
\usepackage{makecell}
\titleformat{\section}{\normalfont\large\bfseries}{\thesection.}{0.6em}{}
\titleformat{\subsection}{\normalfont\normalsize\bfseries}{\thesubsection.}{0.6em}{}
\titlespacing{\section}{0pt}{1.4ex plus 1ex minus .2ex}{1ex}
\makeatletter
\newcommand{\papertitle}[1]{\gdef\@papertitle{#1}}
\newcommand{\printtitleblock}{%
  \begin{center}
    {\fontsize{17}{20}\selectfont\bfseries \@papertitle}\\[0.55em]
  \end{center}
}
\makeatother

\begin{document}
\thispagestyle{empty}

\papertitle{HPC Modeling of Coupled Elastic-Acoustic Wave Propagation in Biological Media: Numerical Validation}
\printtitleblock

\begin{center}
  {\large\bfseries
    Fawad Ali$^{*}$,~~
    Carlos Garc\'ia,~~
    Lapo Boschi
  }\\[0.8em]
  {\small
    Department of Geosciences, University of Padova\\
    Via Gradenigo 6, 35131 Padova, Italy\\
  }
{\footnotesize $^{*}$Corresponding author: \texttt{fawad.ali@unipd.it}}

\end{center}

\vspace{0.8em}
\noindent

\vspace{1em}
\vspace{2.4em}
\begin{center}
{\bfseries\large Abstract}
\end{center}
\noindent
\begin{minipage}{\linewidth}
\small
Accurate numerical models of sound propagation through biological media are an important tool for many applications, from medical physics to studying the auditory system of humans or other animals. We model high-frequency elastic and acoustic wave propagation through the head anatomy of a common bottlenose dolphin (Tursiops truncatus), by means of the open-source software packages \textsc{specfem3d}, based on the spectral-element method, and \textsc{{\it k}-wave}, based on the pseudospectral method. To achieve sufficiently high performance, we ported the latter solver to C++, for Multi-GPU CUDA support via the slab-decomposition of three-dimensional Fast Fourier Transform (3D FFT) approach. Because the two schemes differ fundamentally in how the propagation medium is discretized and internal (in particular fluid-solid) interfaces are treated, similarity between modeled signals across methods is a legitimate measure of model accuracy. Plane waves, depending on time like four-cycle sinusoidal bursts of varying central frequency (between 20 and 100 kHz), are numerically propagated through a computed-tomography-based anatomy model using both solvers. The sound is ``recorded'' in front of the rostrum and at the right inner-ear locations for comparison. We successfully cross-validate both solvers on high-fidelity High-Performance Computing (HPC) clusters. We find that the stability of results from both solvers grows as the corresponding spatial resolution is refined. At their highest resolutions, the two methods show excellent agreement, with normalized correlation exceeding 0.99 across the entire 20–100 kHz frequency range. Together, our results provide a validated HPC simulation framework for wave propagation in biological media, with broader implications, e.g., for biosonar research, auditory biomechanics, and medical ultrasound.
\end{minipage}

\vspace{1.8em}
\vspace{1.8em}
\noindent \rule{\linewidth}{0.4pt}
\vspace{2em}

\section{Introduction}
\noindent In recent years, methods from seismology and applied geophysics have increasingly been applied to topics in medical physics, for example to develop acoustics-based, high-resolution imaging techniques \citep{bachmann2020source,guasch2020full,marty2024transcranial}. In two recent studies \citep{ali2025pseudo, garcia2026feasibility}, our team has adapted numerical wave propagation models to one specific problem in bioacoustics: understanding the contribution of various anatomical features to sound (echo)localization by toothed whales, known to achieve extremely high accuracy despite the lack of external ears \citep{renaud1976sound,au1993sonar,de2023neuroanatomy,ketten1997structure,houser2025marine,reinwald2018bone,hejazi2020contribution}. 
Numerical modeling is particularly useful in this context, because, while research in human audition has relied on databases of head-related impulse responses/transfer functions (HRIRs/HRTFs) recorded directly near the eardrum of subjects\citep{algazi2001}, the vestigial ear canals of dolphins are occluded, making the inner ear inaccessible, and direct measurements impossible. 

Theories of toothed-whale sound reception \citep[e.g.,][]{norris1968,aroyan2001three,cranford2008acoustic,cranford2010, zhang2017directional,wei2018numerical,wei2018finite,wei2020modeling,wei2024validated} have often been tested via numerical models. Many studies, however, have relied on simplified, 2D wave-propagation solvers. Most of the few available 3D models are ``frequency-domain'', i.e., wave propagation is modeled independently at individual frequencies; calculating and visualizing its time dependence is cumbersome and time-consuming. This makes it difficult to analyze and interpret the multiple reflections which are likely to play a fundamental role in source localization\citep{hejazi2020contribution}. Fully 3D time-domain simulations of wave propagation in realistic biological media are computationally demanding. For instance, while soft tissues exhibit acoustic impedances close to that of water (and are often treated as fluids \citep{marty2021acoustoelastic}), bone tissues are elastic solids, with a full stress tensor to be modeled together with the velocity field: the sharp contrast between scalar acoustic and elastic media introduces significant computational overhead, and this effect is further amplified at the ultrasonic frequencies typical of dolphin biosonar,
with the required spatial resolution becoming extremely fine.

Several numerical methods are available for simulating wave propagation in complex biological media\citep[][for a review]{igel2017computational}, including the finite-element method (FEM), the spectral-element method (SEM), and pseudospectral method (PSM). As this proposal is being written, besides our own work\citep{ali2025pseudo,garcia2026feasibility}, we are aware of only another published study in recent literature implementing fully 3D time-domain simulation of wave propagation through a cetacean head\citep{wei2024validated}. Taken together, these efforts suggest that \textsc{COMSOL}, \textsc{SPECFEM3D} and  \textsc{{\it k}-wave} (based on FEM, SEM and PSM solvers, respectively) are possibly the best currently available software packages for our goals. Those studies are limited, however, to frequency $< 50$ kHz, while the central frequency of cetacean echolocation clicks is closer to $\sim 100$ KHz. We are currently striving to push the current limits much further, modelling fully 3D wave propagation models up to at least $\sim 150$KHz. To achieve this, it is crucial that we assess their relative effectiveness and identify the best-suited one(s): this is one of the objectives of the present contribution. 

Despite the maturity of all these methods, we are not aware of a detailed benchmark in the context of broadband wave propagation through complex biological media. Such a comparison is particularly relevant given their contrasting strengths/weaknesses: for instance, (i) SEM honours fluid-solid interface conditions but requires high-quality hexahedral meshes, whereas (ii) PSM operates on {regular Cartesian} grids but may suffer from ``staircase-effect'' artifacts \citep{van2002finite,ali2025pseudo} at material discontinuities. In this work, we present a systematic benchmark between SEM and PSM, including our Compute Unified Device Architecture (CUDA) version of the latter (hereafter referred to as \textsc{psm-cuda}), applied to wave propagation through a three-dimensional dolphin's head model at central frequencies ranging from 20 to 100 kHz. We assess numerical accuracy, convergence and computational cost on high-performance computing clusters. 

\section{Summary of the Pseudo-Spectral Method (PSM)}

\noindent We make use of the \textsc{{\it k}-wave} open-source software, currently available via \url{http://www.k-wave.org/}, with some technical adaptations described in the final portion of this section.
\textsc{{\it k}-wave} was originally conceived as a ``toolbox for \textsc{matlab}'', but the distribution now includes an ``optimised C++ version of the code that maximises computational performance for large simulations''. \textsc{{\it k}-wave} has a number of features that make it particularly interesting for our application. It is ``designed for time domain acoustic and ultrasound simulations in complex and tissue-realistic media''. It honours both acoustic and elastic waves\citep{treeby14}. In earlier work by our team, \textsc{{\it k}-wave} was already used to model wave propagation through the {\it skull} of a dolphin, neglecting the effects of soft tissues\citep{hejazi2020contribution}.

\subsection{The \textsc{{\it k}-wave} approach: main ideas}\label{sec:k-wave-basics}

The $k$ in \textsc{{\it k}-wave} refers to the fact that, before implementing a finite-difference (FD) -type scheme to calculate the solution at time $t+\delta t$ from the solution at $t$, the equation of motion is {spatial-Fourier-transformed}. (And $k$ usually denotes spatial frequency, just like $\omega$, or $\nu$, usually denote frequency with respect to time.) 

In a basic FD scheme, one discretizes the derivatives of the unknown function with respect to time and space; e.g., it can be proved, via the Taylor formula, that 
\begin{equation}\label{disc_2d_der}
\frac{d^2f}{dx^2}(x_i) \approx \frac{f(x_{i+1}) + f(x_{i-1}) - 2 f(x_i)}{\delta x^2},
\end{equation}
is fourth-order accurate in $\delta x$ \citep[e.g.,][]{igel2017computational}. For instance, in an inviscid fluid, {\it pressure} $p$ obeys the wave equation, so we use eq. (\ref{disc_2d_der}) to discretize the second derivatives that appear in it, with respect to time and distance, and end up with an {\it algebraic} (``difference'') equation, which we can solve for the value of $p$ at time $t_{k+1}$, to find that (in the simplest, 1-D case)
\begin{equation}\label{N9_iii2}
p(x_{i},t_{k+1}) = c^2 \frac{\delta t^2}{\delta x^2} \left[ p(x_{i+1},t_k) + p(x_{i-1},t_k) - 2 p(x_i,t_k) \right] + 2 p(x_i,t_k) -  p(x_{i},t_{k-1}),
\end{equation}
where $c$ denotes wave speed, and the meaning of all other symbols should be intuitive.

Eq. (\ref{N9_iii2}) is pure FD. An alternative approach consists of expressing the unknown function ($p$ in the example above) as the linear combination of a set of known {\it basis functions} (before discretization). That is called a {\it spectral}, or {\it pseudospectral} method. In the case of $k$-wave, specifically, $p(x_1, x_2, x_3, t)$ is written as a linear combination of $\sin(k_1 x_1)$, $\sin(k_2 x_2)$, $\sin(k_3 x_3)$, $\cos(k_1 x_1)$, $\cos(k_2 x_2)$, $\cos(k_3 x_3)$, 
through coefficients that are still dependent on $t$. In other words, $p$ is {\it spatial-Fourier-transformed}. This is described \citep[equation (3) of][]{tabei02} by
\begin{equation}\label{tabei3}
\frac{\partial ^2 p({\bf k},t)}{\partial t^2} = -c^2 |{\bf k}|^2 p({\bf k},t),
\end{equation}
where ${\bf k}=(k_1, k_2, k_3)$,
and $p({\bf k},t)$ is the spatial-Fourier-transform of $p({\bf x},t)$. 

In practice, the Fourier transform is really a {\it discrete} Fourier transform, i.e. 
$k_1$, $k_2$, $k_3$ can only take a discrete set of values. That is how spatial dependence is discretized, which is fundamentally different from the FD approach. In time, eq. (\ref{tabei3}) can be discretized as follows\citep{tabei02},
\begin{equation}\label{tabei5}
\frac{c^2 |{\bf k}|^2 \delta t^2}{4 \sin^2\left(\frac{c |{\bf k}| \delta t}{2}\right)}
\frac 
{p({\bf k},t+\delta t) + p({\bf k},t-\delta t) - 2 p({\bf k},t)}
{\delta t^2} 
= -c^2 |{\bf k}|^2 p({\bf k},t),
\end{equation}
which can be solved for $p({\bf k},t+\delta t)$, similar to a FD scheme. 
Eq. (\ref{tabei5}), which is often referred to as the ``harmonic-oscillator differential equation'', is reminiscent of (\ref{disc_2d_der}), i.e., of the FD discretization, but clearly it is not the same thing. The proof of eq. (\ref{tabei5}) is given in appendix A.

Importantly, the difference equation (\ref{diffeq2}) does not result from an approximation, but is exact: which contributes to the accuracy of this scheme and of the $k$-space approach\citep{tabei02}.

While the purpose of the previous paragraphs was to illustrate the central ideas behind \textsc{{\it k}-wave} and the PSM, the \textsc{{\it k}-wave} toolbox includes a number of technically important features that we have not yet mentioned. In particular, the actual  \textsc{{\it k}-wave} algorithm involves a ``staggered grid''/``leapfrog scheme'' \citep[e.g.,][]{igel2017computational}, which contributes to its accuracy\citep{tabei02}, absorbing boundary conditions are implemented via a ``perfectly matched layer'' (i.e., ``a thin absorbing layer that encloses the computational domain and is governed by a nonphysical set of equations, causing anisotropic attenuation''\citep{treeby10}).

\subsection{\textsc{{\it k}-wave} implementation of the elastic equations}

In \textsc{{\it k}-wave}, the implementation of wave propagation in {\it elastic} (as opposed to purely acoustic) media\citep{treeby2014modelling} is based on the stress-velocity formulation of the momentum equation, i.e.\citep{virieux1984sh}, 

\begin{flalign}
  &\rho(\mathbf{x})\,\partial_t v_x
= \partial_x \sigma_{xx}   + \partial_y \sigma_{xy}+ \partial_z \sigma_{xz}, \label{eq:mom-x} \\[2pt]
 &\rho(\mathbf{x})\,\partial_t v_y
 = \partial_x \sigma_{xy}
   + \partial_y \sigma_{yy}
   + \partial_z \sigma_{yz}, \\[2pt]
 &\rho(\mathbf{x})\,\partial_t v_z
 = \partial_x \sigma_{xz}
   + \partial_y \sigma_{yz}
   + \partial_z \sigma_{zz}, \\[2pt]
  &\partial_t \sigma_{xx}
= (\lambda(\mathbf{x})+2\mu(\mathbf{x}))\,\partial_x v_x
   + \lambda(\mathbf{x})\,(\partial_y v_y + \partial_z v_z), \\[2pt]
&\partial_t \sigma_{yy}
  = (\lambda(\mathbf{x})+2\mu(\mathbf{x}))\,\partial_y v_y
   + \lambda(\mathbf{x})\,(\partial_x v_x + \partial_z v_z), \\[2pt]
 &\partial_t \sigma_{zz}
 = (\lambda(\mathbf{x})+2\mu(\mathbf{x}))\,\partial_z v_z
   + \lambda(\mathbf{x})\,(\partial_x v_x + \partial_y v_y), \\[2pt]
&\partial_t \sigma_{xy}
  = \mu(\mathbf{x})\,(\partial_y v_x + \partial_x v_y), \\[2pt]
 &\partial_t \sigma_{xz}
 = \mu(\mathbf{x})\,(\partial_z v_x + \partial_x v_z), \\[2pt]
&\partial_t \sigma_{yz}
  = \mu(\mathbf{x})\,(\partial_z v_y + \partial_y v_z), \label{eq:vel}
\end{flalign}
\noindent {valid for an isotropic lossless medium.}
No explicit interface condition is prescribed at boundaries contained within the medium of propagation (e.g., fluid-solid boundaries).
The only distinction between elastic and acoustic regions ($\Omega_s$ and $\Omega_f$, respectively) lies in the values of the elastic parameters $\rho(\mathbf{x})$,
$\lambda(\mathbf{x})$, and $\mu(\mathbf{x})$, 
or their relations to acoustic and compressional and shear wave speeds ($c_a, V_P$ and $V_S$, respectively), i.e., 
\begin{equation}
\lambda(\mathbf{x}) =
\begin{cases}
\rho_s(\mathbf{x})\bigl[V_P^2(\mathbf{x})
  - 2V_S^2(\mathbf{x})\bigr] & \mathbf{x} \in \Omega_s, \\[3pt]
\rho_f(\mathbf{x})\,c_a^2(\mathbf{x})
  & \mathbf{x} \in \Omega_f,
\end{cases}
\qquad
\mu(\mathbf{x}) =
\begin{cases}
\rho_s(\mathbf{x})\,V_S^2(\mathbf{x})
  & \mathbf{x} \in \Omega_s, \\[3pt]
0 & \mathbf{x} \in \Omega_f.
\end{cases}
\end{equation}
Setting $\mu = 0$ in $\Omega_f$ eliminates all shear stress
components, collapsing the stiffness tensor to
the acoustic bulk modulus $\kappa_a = \lambda = \rho_f c_a^2$.
The traction and kinematic continuity conditions at the interfaces between
$\Omega_s$ and $\Omega_f$
are not enforced explicitly; they are approximated implicitly
through the jump in material properties across the grid cell
containing the interface, introducing an $\mathcal{O}(h)$ error
initially localised to that cell, but that might in general propagate across the medium \citep[e.g.,][]{van2002finite}.

\subsection{Leapfrog scheme}

Eqs. (\ref{eq:mom-x})-(\ref{eq:vel}) are solved according to the basic idea described in sec. \ref{sec:k-wave-basics}; the algebra is of course much more complicated as we are now in 3D and in an elastic, rather than acoustic medium. Let us summarize the main features of the algorithm.

As anticipated, all spatial derivatives are computed globally via spatial Fast Fourier Transform (FFT). 
For a 3D field $f(\mathbf{x})$ on a Cartesian grid, with uniform spacing $h$,
\begin{equation}
\partial_\alpha f \;=\;
\mathcal{F}^{-1}\!\left\{
  ik_\alpha^{\pm}\;\mathcal{F}\{f\}(\mathbf{k})
\right\}, \qquad \alpha \in \{x,\, y,\, z\},
\end{equation}
where $\mathcal{F}\{f\}$ denotes the FFT of $\{f\}$, $i$ the imaginary unit, and the so called  ``staggered wavenumber operators''
\begin{equation}
k_\alpha^{\pm}
= k_\alpha\,e^{\pm ik_\alpha h/2}
\end{equation}
shift the evaluation point by $\pm h/2$ along {the coordinate} $\alpha$
to match the staggered node positions of velocity vector $\mathbf{v=v_x,v_y,v_z}$
(at half-nodes) and the stress tensor $\boldsymbol{\sigma}$ (at integer nodes).

The velocity components $\mathbf{v}$ live at half time-steps
$n \pm \tfrac{1}{2}$ and the stress components
$\boldsymbol{\sigma}$ at integer steps $n$
(staggered leapfrog):
\begin{align}
\mathbf{v}^{n+1/2}
  &= \mathbf{v}^{n-1/2}
   + \frac{dt}{\rho}\left(
       \mathcal{F}^{-1}\!\left\{
         i\mathbf{k}^{+}\,\mathcal{F}\{\boldsymbol{\sigma}^n\}
       \right\}
       + \mathbf{f}^n
     \right), \\[8pt]
\boldsymbol{\sigma}^{n+1}
  &= \boldsymbol{\sigma}^{n}
   + dt\;\mathbf{C}\!:\!
     \mathcal{F}^{-1}\!\left\{
       i\mathbf{k}^{-}\,\mathcal{F}\{\mathbf{v}^{n+1/2}\}
     \right\},
\end{align}
where $i\mathbf{k}^{\pm}$ denotes the appropriate staggered
wavenumber operator for each derivative component.

\subsection{Optimizations and adaptations of \textsc{{\it k}-wave} software}

\subsubsection{Single-GPU implementation}
In practice, the most straightforward way to implement the {\it elastic} equations (\ref{eq:mom-x}) through (\ref{eq:vel}) in \textsc{{\it k}-wave}  is via the \texttt{pstdElastic3D.m} \textsc{matlab} program\cite{treeby2014modelling}.

Unlike the {\it acoustic} solver of the \textsc{{\it k}-wave} toolbox \cite{treeby2010k}, which applies a \textit{k}-space temporal correction to the time-marching scheme, \texttt{pstdElastic3D.m} relies on a staggered-grid leapfrog scheme, and therefore requires a smaller Courant–Friedrichs–Lewy (CFL) number\citep{lewy1928partiellen} to maintain accuracy and stability. The smaller the CFL number, the shorter the time step needed to achieve a certain spatial resolution of the wave field, the higher the computational cost. We kept a CFL value below 0.289 throughout the simulations. As explained above, simulating ultrasound propagation through dolphin anatomy requires that even high-performance computational resources be pushed to their current limits. Accordingly, we were careful to use \textsc{{\it k}-wave} in such a way as to maximimze its performance.  

The baseline implementation of \texttt{pstdElastic3D.m} targets CPU execution; however, by performing the FFT operations on GPU via \textsc{matlab}'s \texttt{gpuArray} interface, it is possible to achieve Single-GPU acceleration with minimal code modification. For simulation domains small enough to fit within the GPU video memory (VRAM), the single-GPU version is significantly more performant than the CPU one. 

\subsubsection{Multiple-GPU implementation}
The major drawback of running \textsc{{\it k}-wave} on a single GPU is that memory usage cannot exceed the VRAM capacity of a single GPU. In the PSM, on the other hand, spatial differentiation via FFT requires the entire spatial domain to be available as a contiguous array. 
In our case, available single-GPU VRAM was exceeded for simulations of grid spacing below 0.9 mm. 

To overcome this limitation, we re-implemented the \texttt{pstdElastic3D.m} algorithm in CUDA C++ for lossless elastic media, with full Multi-GPU support. In practice, this allows us to access the entire 640 GB VRAM of up to 8 H100 GPUs on a single DGX (Deep GPU Xceleration) node (on the hardware available to us), i.e. one order-of-magnitude more VRAM than on a single GPU. 

In our implementation, the arrays where we store medium-parameter and initial-condition values  are exported from \textsc{matlab} as binary files and loaded directly by the CUDA code. All GPU kernels are written in CUDA and operate on single-precision floating-point arrays residing in device memory. Because the VRAM allowed by the multi-GPU approach is very large, {2D domain decomposition} (also known as pencil-decomposition) for parallel 3D FFT \citep{li20102decomp,ayala2013parallel,fang2007performance,romero2022distributed} is not strictly necessary; adapting the common {1D domain decomposition} (or slab-decomposition) approach \citep{dmitruk2001scalable} was enough to achieve our goals.

The global domain $\Omega$, of size $N_x \times N_y \times N_z$, is partitioned into $N_\text{GPU}$ slabs along the $z$-direction, each of size $N_x \times N_y \times N_z^{\text{loc}}$, where $N_z^{\text{loc}} = N_z / N_\text{GPU}$. This decomposition has the property that the full extent of the domain in $x$ and $y$ is preserved within each GPU, so FFTs along $x$ and $y$ can be performed independently and locally on each device without inter-GPU communication.

The $z$-direction FFT, however, requires data from all slabs simultaneously. To perform it, a global ``pencil transpose'' is executed: the domain is redistributed so that each GPU holds a contiguous block in $x$ while the full $z$-extent is local. Concretely, after the transpose each GPU owns a pencil of size $N_x^{\text{loc}} \times N_y \times N_z$, where $N_x^{\text{loc}} = N_x / N_\text{GPU}$. The $z$-FFT is then performed locally on each GPU. The transpose and the associated data exchange between GPUs are implemented using NVIDIA Collective Communication Library (NCCL) all-to-all collectives over NVLink, which provides the high-bandwidth peer-to-peer interconnect required for this operation. After the $z$-FFT the inverse transpose restores the slab layout, and the simulation proceeds to the next time step.

Except where otherwise specified, all PSM simulations presented below were carried out with the multi-GPU \textsc{psm-cuda} approach. Compared with the Single-GPU implementation, the Multi-GPU version introduces two additional global communication steps per time step (forward and inverse transpose), but scales the available VRAM linearly with the number of GPUs, enabling simulations on domains that would otherwise be insufficient.

\subsubsection{ \textsc{{\it k}-wave} source implementation}
\textsc{{\it k}-wave} requires that forcing be prescribed in the form of a set of vectors, one per grid node where forcing is applied, and one coefficient per time step, extending over the entire temporal extent of a simulation. Since we have chosen to model {\it plane} waves, this requires that a large number of such vectors are stored in VRAM, quickly filling up to several GBs and resulting in a significant bottleneck. 

To alleviate VRAM usage, we subdivide each full simulation, involving a very large number of time steps, into a sequence of shorter simulations (``stages''), each running only for a fraction of the total number of time steps. The stages are run sequentially, and the velocity and stress fields obtained at the end of each are saved to disk, then prescribed as the initial conditions (the ``source'') of the next. For example, a simulation of the full dolphin head model (section~\ref{sec:dolphin_head_model} below, $42 \times 42 \times 65$ cm), with 0.8~mm grid spacing, would require approximately 13.5~GB of extra VRAM to store the source matrix on a single GPU, however, subdividing the simulation into 10 stages reduces the peak VRAM requirement by approximately 12.15~GB.
\section{Summary of the Spectral-Element Method (SEM)}

\subsection{Hexahedral mesh}\label{sec:carlosmesh}

\noindent Throughout this study, SEM is implemented via \textsc{specfem3d cartesian} (hereafter simply referred to as \textsc{specfem3d}) \citep{komatitsch2002spectral,komatitsch2002spectral2}, a widely used open-source solver for seismic wave propagation.

Simply put, SEM is a form of FEM, where the basis functions used to discretize the medium of wave propagation are selected so as to make the time stepping particularly fast, and the RAM requirements particularly low. In practice, both SEM and FEM involve the inversion of a so-called ``mass matrix'' at each time step (think a much more complex version of the 1D, non-staggered FD eq. (\ref{N9_iii2}) above), but, by its clever parameterization, the mass matrix of SEM is diagonal (while that of FEM is not, resulting in a significant bottleneck)\citep[e.g.,][ch. 6 and 7]{igel2017computational}. The SEM parameterization consists of discretizing the medium on a grid (``mesh'') of {\it hexahedra}, as opposed to the {\it tetrahedra} typical of FEM, and then parameterizing the hexahedra 
\begin{center}
    \includegraphics[width=0.7\linewidth]{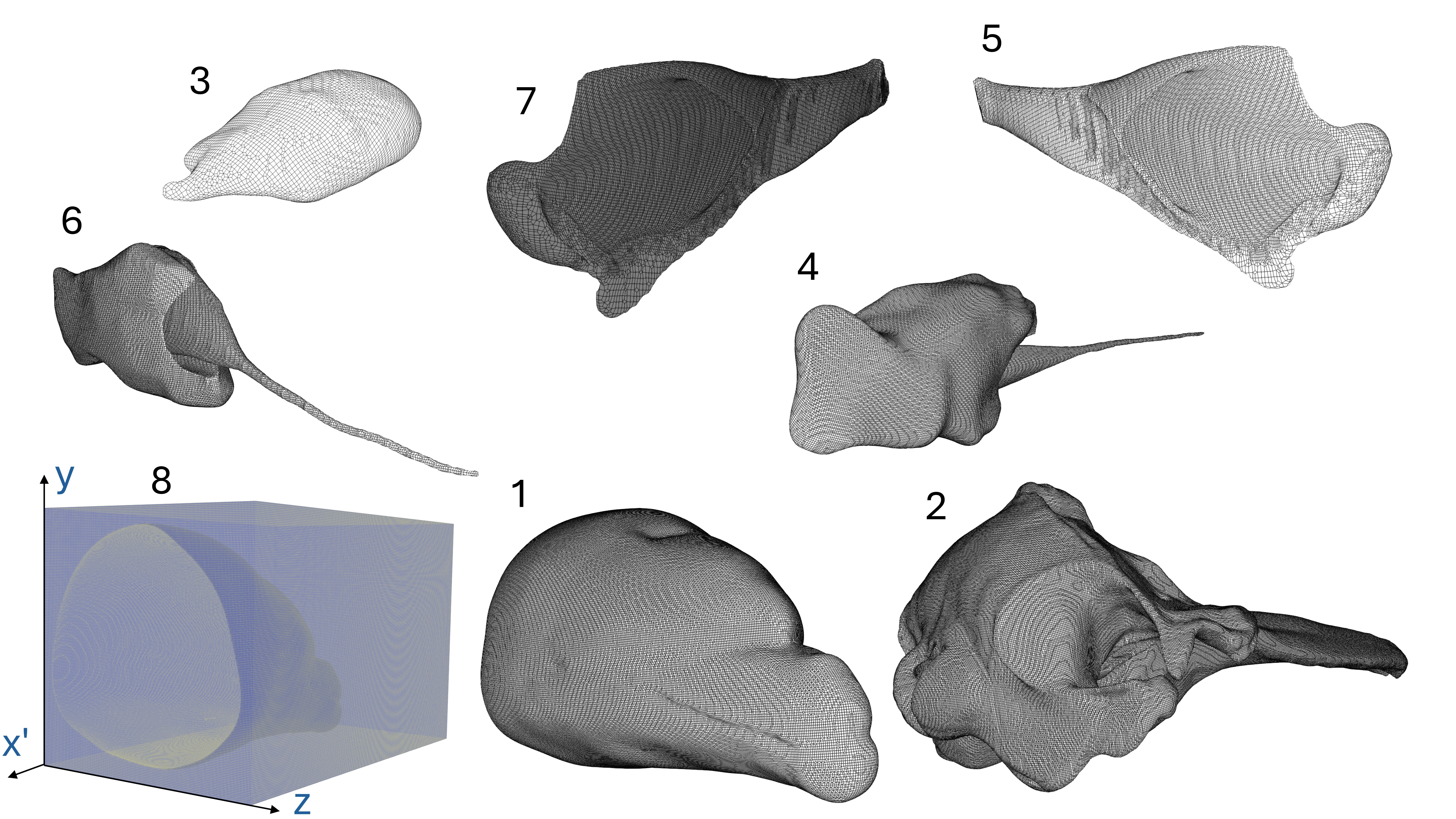}\label{fig:mesh}
    \captionof{figure}{\textit{Tursiops truncatus} mesh\citep{garcia2026feasibility}, showing the eight homogeneous regions into which the head model is subdivided:
    (1) soft tissue, (2) skull bone, (3) melon, (4) acoustic fats (left side),
    (5) lower jaw bone (right side), (6) acoustic fats (right side),
    (7) lower jaw bone (left side), (8) water.}
    \label{fig:dolphin_mesh}
\end{center}

\noindent themselves in terms of one set of Lagrange polynomials per spatial coordinate. The number of such polynomials can be increased (decreased) to enhance (diminish) resolution, with higher (lower) computational costs; it coincides with the number of ``Gauss-Lobatto-Legendre'' (GLL) points, or $NGLL$, i.e. the few discrete points where the continuous function to be discretized will coincide exactly with its discrete form\citep[e.g.,][ch. 6 and 7]{igel2017computational}.

Unlike PSM, both FEM and SEM meshes are ``conformal'', i.e. physical discontinuities approximately coincide with boundaries between neighboring hexahedra. As a result, hexahedra can be significantly distorted with respect to regular voxels; a certain amount of distortion is necessary; a lot of distortion can result in significant numerical noise, which must be avoided. Distortion is quantified by the ``scaled Jacobian'' of a given hexahedron,

$$sj=\frac{\det (A)}{|a_1|\,|a_2|\,|a_3|},$$ 

\noindent where $A = [ a_1,  a_2,  a_3]$ , and $a_1$, $a_2$, and $a_3$ are the three edge vectors drawn from a reference node to its three adjacent nodes\citep{knupp2000achieving}. In the absence of distortion, i.e. when the hexahedron is a perfect cube, $sj=1$. While most authors agree, to our knowledge, that values of $sj$ lower than 0.2 should be avoided\citep{garcia2026feasibility}, and values of $sj$ above 0.5 are relatively safe, there is no rigorous way to tell a whether a certain level of mesh distortion will result in significant numerical noise. This is one of the reasons for the cross-validation exercise that we present here. 

In earlier work by our team, computed-tomography (CT) -based scans of a deceased toothed whale's (\textit{Tursiops truncatus}) head, provided by colleagues at the Department of Comparative Biomedicine and Food Science, University of Padua, were meshed on a hexahedral grid (shown here in Fig.~\ref{fig:dolphin_mesh}). For the sake of simplicity, after identifying eight approximately homogeneous regions, the values of all elastic parameters ($\rho$, $V_P$, $V_S$) were chosen to be constant in each region (each parameter coinciding to its average within each region). A few \textsc{specfem3d} simulations were run on the resulting model\citep{garcia2026feasibility}. In the following, we shall present the results of a suite of additional simulations conducted on the exact same mesh, with a range of different values of $NGLL$ and different frequency spectra for the source signal.

The complete mesh consists of 12{,}676{,}714 hexahedra, with statistics for each region reported in Table~\ref{tab:mesh_stats}. The values of minimum and maximum edge length, $h_{\min}$ and $h_{\max}$, are particularly relevant as (together with $NGLL$) they control the Courant-Friedrichs-Lewy (CFL) stability condition\citep{lewy1928partiellen} (and time-step length), and the highest frequency that can be adequately modelled with a given mesh. While Table~\ref{tab:mesh_stats} shows that the mesh includes a wide range of element sizes, most hexahedra are relatively regularly shaped, with the overall average $sj$ equal to 0.9207, and about 90\% of all hexahedra throughout the mesh having a maximum edge length of exactly 2.519\,mm. 
\begin{table}[!htbp]
  \centering
  \fontfamily{ptm}\selectfont
  \renewcommand{\sfdefault}{ptm}
  \caption{Mesh quality and h (element-size)  statistics per region of the
           hexahedral Cubit mesh. The scaled Jacobian \emph{sj} is dimensionless;
           element sizes are in metres. The $h_{90\%}$ measures the value of $h_{\max}$ below which
90\% of the cells fall.}
  \label{tab:mesh_stats}
  \footnotesize
  \setlength{\tabcolsep}{7pt}
  \renewcommand{\arraystretch}{1.10}
  \begin{tabular}{lrrcccccc}
    \toprule
    Region & Elements & Nodes
           & $sj_{\min}$ & $sj_{\text{mean}}$
           & h$_{\min}$ & h$_{\max}$ & h$_{\text{mean}}$ & h$_{90\%}$ \\
    \midrule
    1 & 3{,}768{,}014 & 4{,}106{,}132 & 0.1835 & 0.8357 & 0.000529  & 0.004848 & 0.002126 & 0.002768 \\
    2 & 1{,}952{,}293 & 2{,}156{,}614 & 0.1928 & 0.8995 & 0.0004455 & 0.002594 & 0.001299 & 0.001493 \\
    3 &    48{,}804   &    53{,}796   & 0.3082 & 0.9093 & 0.0009371 & 0.003924 & 0.002564 & 0.002801 \\
    4 &   557{,}884   &   601{,}285   & 0.1878 & 0.9291 & 0.0004965 & 0.002634 & 0.001286 & 0.001446 \\
    5 &   138{,}782   &   162{,}400   & 0.1571 & 0.7801 & 0.0005012 & 0.004848 & 0.001784 & 0.002625 \\
    6 &   525{,}844   &   566{,}375   & 0.1811 & 0.9311 & 0.0006024 & 0.002534 & 0.001286 & 0.001451 \\
    7 &   137{,}240   &   159{,}854   & 0.1966 & 0.7856 & 0.0005007 & 0.004053 & 0.001813 & 0.002666 \\
    8 & 5{,}547{,}853 & 5{,}711{,}978 & 0.302  & 0.991  & 0.0006826 & 0.003838 & 0.002509 & 0.0025   \\
    \midrule
    \textbf{All regions}
      & \textbf{12{,}676{,}714} & \textbf{12{,}795{,}549}
      & \textbf{0.1571} & \textbf{0.9207}
      & \textbf{0.0004455} & \textbf{0.004848}
      & \textbf{0.002089} & \textbf{0.002519} \\
    \bottomrule
  \end{tabular}
  \footnotesize
\end{table}

\subsection{Interface and boundary conditions}
All SEM simulations were conducted using the open-source distribution of \textsc{specfem3d}, compiled and run in several configurations, including GPU mode, CPU mode, and different $NGLL$  values, but with no modifications on our part. We refer the readers to the existing literature on SEM and  \textsc{specfem3d} for details on how the algorithm exactly works\citep[e.g.,][]{komatitsch2002spectral,komatitsch2002spectral2}. 

Importantly, unlike PSM, interface conditions in \textsc{specfem3d} are prescribed explicitly at all interfaces embedded within the medium, and in particular at fluid-solid discontinuities, where numerical errors would be likely to arise otherwise\citep[e.g.,][]{van2002finite}. Namely, (i) scalar pressure on the fluid side is required to equal the component of traction normal to the interface on the solid side, and (ii) the component of velocity normal to the interface is required to be continuous across the interface. 

Finally, \textsc{specfem3d} allows for two different implementations of absorbing boundary conditions at the outer edges of the simulation domain, i.e. Convolutional Perfectly Matched Layer (CPML)\citep{komatitsch2007unsplit} and ``Stacey''\citep{clayton1977absorbing,stacey1988improved}. The former employs the unsplit version of PML\citep{berenger1994perfectly}, by incorporating complex stretching coordinates to enhance the absorption of outgoing waves, whereas the latter is a zero-thickness absorbing boundary condition based on the paraxial wave equation that suppresses reflections at the domain edges. Importantly, the CPML option is not supported in the GPU build of \textsc{specfem3d}: after experimenting with both approaches and verifying their consistency, we eventually applied Stacey boundary conditions to most of the simulations presented here, as discussed in the following.
\section{Benchmark Protocol}\label{subsec:benchmarking}

\noindent We next describe our procedure for ``benchmarking'', i.e., cross-validating the chosen implementations of PSM and SEM, independent of the propagation medium/model. In practice, this is repeated here for two models, a very simple one (two concentric spheres, made of bone and soft tissue and immersed in water) and a very complex one (the dolphin head, also immersed in water), as described later in sec.~\ref{sec:results}. 

The {\it hexahedral} mesh originally prepared for \textsc{specfem3d} is the common starting point for both solvers, to make sure that we shall be comparing wave propagation simulation results rather than the effects of discretizing CT scans in different ways.

The mesh is formatted for \textsc{specfem3d} as a collection of files, including \texttt{nodes\_coords\_file}, which contains the coordinates of the mesh nodes, \texttt{mesh\_file}, which defines the element connectivity, and additional files specifying material properties and boundary conditions. A \textsc{python} script assembles these files and converts the mesh into a Visualization Toolkit (VTK) file. This file is then imported into \textsc{paraview}, where the \texttt{Resample to Image} filter converts the mesh to a regular Cartesian grid. The resulting data are exported as a \texttt{.mat} file for further processing in \textsc{matlab}.

Hexahedral and Cartesian meshes of both benchmark models are compared in 
Figure~\ref{fig:dolphin_workflow}. At a voxel resolution of 0.8~mm, the soft-tissue regions show good visual agreement with the hexahedral mesh (Figure~\ref{fig:dolphin_workflow}e--f); however, zooming in reveals the \textsc{{\it k}-wave} pixelization inherent to rectilinear voxel grids at curved boundaries. {This is likely to introduce an error in wave propagation simulations, that can be reduced by enhancing grid resolution \citep{van2002finite,ali2025pseudo}. One goal of our study is to verify that this error becomes negligible with increasing resolution.}

\subsection{Implementation of the source}\label{sec:source}
While straightforward in principle, implementation of the source of wave propagation is a very delicate step of the benchmark procedure, because of the limitations of both \textsc{{\it k}-wave} and \textsc{specfem3d} in how a source can be represented, and the need to excite both \textsc{{\it k}-wave} and \textsc{specfem3d} simulations with the exact same source. In practice, both \textsc{specfem3d} and \textsc{{\it k}-wave} are excited by combinations of point sources, each with its prescribed source time function. In the case of \textsc{specfem3d} this means that, as long as the source is within the fluid medium (which is the case throughout this study), the scalar potential from which both pressure and displacement/velocity/acceleration are derived is perturbed at the source

\begin{center}
    \includegraphics[width=\linewidth]{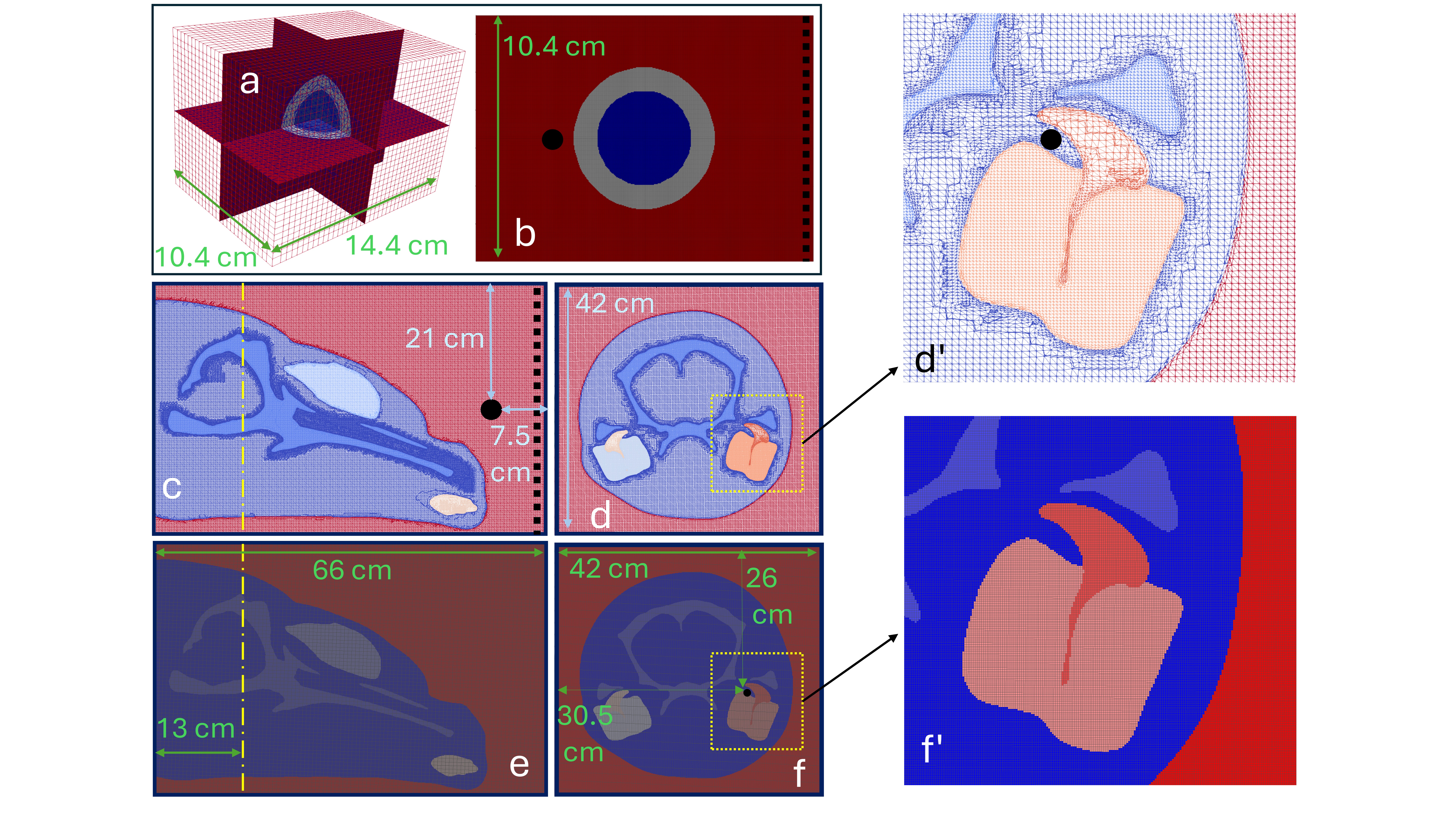}
    \captionof{figure}{Workflow for constructing consistent \textsc{{\it k}-wave} and \textsc{specfem3d}, i.e., conformal hexahedral and Cartesian parameterizations for benchmarking. Point sources are deployed along the black dotted line (corresponding to a plane in 3D), while two black dots denote sensor locations of two and the black dots represent the sensor locations in both sphere (panels a,b) and dolphin-head (panels c through f') models. (a) simple hexahedral mesh of two concentric spheres, and (b) 2D section of the corresponding regular-grid model (input of the PSM) obtained from the mesh; (c) mid-sagittal section through the dolphin mesh, and (d) another section through the same mesh, perpendicular to the first and cut along the yellow line of panel (c); (e, f) same sections, regular-grid parameterization of the same model, derived from the hexahedral mesh; (d', f') enlarged details of panels (d) and (f).}
\label{fig:dolphin_workflow}

\end{center}

\noindent location, given by its $\xi$, $\eta$, and $\gamma$ coordinates in the local reference frame of the respective mesh element. In the case of \textsc{{\it k}-wave}, a perturbation to the stress tensor is forced at the corresponding grid nodes. If one then provides the same numerical values as source parameters of  \textsc{specfem3d} and \textsc{{\it k}-wave} simulations, the output of  \textsc{specfem3d} will have to be integrated over time, or the output of \textsc{{\it k}-wave} differentiated with respect to time, before comparison. Throughout this study, the former option was preferred. 

All simulations in this study start with a plane wave entering the simulation domain from the boundary that faces the rostrum, with wavefront exactly parallel to that boundary. In practice, the same time-dependent signal is emitted simultaneously by a set of point sources uniformly and densely distributed over the boundary \citep{ali2025pseudo, garcia2026feasibility}. The spacing between point sources in  \textsc{{\it k}-wave} is set to coincide with the model grid spacing, while in \textsc{specfem3d} it was fixed at 9 mm. (Further reducing the source spacing in \textsc{specfem3d} has led to UCX (Unified Communication X)\citep{shamis2015ucx} errors during the simulation.) In \textsc{specfem3d}, sources are placed at the boundary, 9.9 cm away from the tip of the rostrum, to allow for maximum constructive interference between relatively widely spaced sources; in \textsc{{\it k}-wave}, sources are closer to the rostrum, approximately 0.4 cm away from the boundary of 1.6-cm thick PML region. Based on Huygens-Fresnel's principle \citep{kinsler2000fundamentals,boschi2025structure}, with such source spacing, we can construct a plane wave of frequency up to 166 kHz in \textsc{specfem3d} and 1500 kHz in \textsc{{\it k}-wave} without aliasing.

In time, the emitted signal consists of a 4-cycle sinusoidal burst, with varying central frequency "$f_c$" and tapered via a ``Tukey window''\citep{harris1978use} with $\alpha=0.7$. We padded the water region of the \textsc{{\it k}-wave} grid by several centimeters on the lateral sides i.e., along $y$ and $z$-directions of the Fig. \ref{fig:dolphin_mesh} (region-8). This way, (i) up to 20 layers of PML can be defined entirely within the surrounding water domain, thereby keeping the dolphin head anatomy away from PML, and (ii) as the none-zero amplitude of the point sources along the source plane introduces a discontinuity in PSM, we applied a spatial Tukey window with $\alpha=0.45$ across the source aperture, i.e., along the  $y$ and $z$-directions of the Fig. \ref{fig:dolphin_mesh} in region-8, to smooth the source injection by gradually reducing the source amplitude (almost to zero) toward the lateral edges. In the case of \textsc{specfem3d}, on the other hand, Stacey boundary conditions are prescribed, so the domain was left unmodified.

\subsection{Similarity metrics}

Before comparison, waveforms are resampled by linear interpolation (we linearly interpolate the time-integrated output of \textsc{specfem3d} at \textsc{{\it k}-wave} time steps), and aligned by cross-correlation (the origin of the time axis is not the same in \textsc{specfem3d} vs. \textsc{{\it k}-wave} or \textsc{psm-cuda} simulations).

Given two pressure traces $s_1(n)$ and $s_2(n)$, with $n$ the time-domain sample index,
we define relative (percent) root-mean-square (RMS) difference 
\begin{equation}
   \mathrm{RMS} = 100 \times
        \frac{\mathrm{rms}\,(s_1-s_2)}
             {\tfrac{1}{2}\left[\mathrm{rms}\,(s_1) + \mathrm{rms}\,(s_2)\right]},
\end{equation}
where $\mathrm{rms}(x) = \bigl(N^{-1}\sum_n x^2(n)\bigr)^{1/2}$, 
and normalized correlation
\begin{equation}
    C = \frac{\displaystyle\sum_{n=1}^{N} s_1(n) s_2(n)}
             {\sqrt{\displaystyle\sum_{n=1}^{N} s_1^{2}(n)}\;
              \sqrt{\displaystyle\sum_{n=1}^{N} s_2^{2}(n)}},
\end{equation}

\section{Results and Discussion}\label{sec:results}
\subsection{Concentric spherical heterogeneities}\label{sec:spheretest}

Our first, idealized model consists of a spherical inner core of 2~cm radius, within a concentric shell of 3~cm radius and a $10.4 \times 10.4 \times 14.4$~cm rectangular box (Figure~\ref{fig:dolphin_workflow}a-b). The center of both spheres coincides with the geometrical center of the box. The core is made of bone-like with homogeneous $V_p=2600$ m/s, $V_s=1800$ m/s and $\rho=1700$ kg/m$^3$; the outer shell is made of soft tissue with homogeneous $V_p=1400$ m/s, $V_s=200$ m/s and $\rho=980$ kg/m$^3$. The surrounding volume is filled with water, i.e. $c_a=1500$ m/s, $\rho_a=1000$ kg/m$^3$, no shearing. All outer boundaries are absorbing, as in all simulations presented here. A hexahedral mesh (input of \textsc{specfem3d}) for the model is first derived, then resampled on a 3D Cartesian grid with uniform spacing as input for PSM, as per sec.~\ref{subsec:benchmarking} and Fig.~\ref{fig:dolphin_workflow}.
 
We use this relatively simple and inexpensive setup to test several features of both SEM and PSM  that we shall later take for granted. In particular, we verify that (i) \textsc{specfem3d} simulation results are not affected by the choice of absorbing boundary condition implementations (Stacey vs. CPML); 
(ii) the \textsc{matlab}, single-GPU \textsc{psm-cuda} and multi-GPU \textsc{psm-cuda} versions of \textsc{{\it k}-wave} give practically coincident results; (iii) \textsc{specfem3d} and \textsc{{\it k}-wave} simulation results are also essentially coincident at this level of complexity. 

All these tests are successful, as summarized by Fig.~\ref{fig:sphere_com}.
All traces in Fig.~\ref{fig:sphere_com} are ``recorded'' at a receiver positioned within the water domain, along the straight line that goes through the spheres' center and is parallel to the 14.4-cm long edges of the box, 3.5 cm away from the spheres' center. All models are excited by
a plane-wave as described in sec.~\ref{sec:source}, entering the box from the boundary antipodal to the receiver, with $f_c=20$ (Fig.~\ref{fig:sphere_com}a) or $80$ (Fig.~\ref{fig:sphere_com}b) kHz. Uniform Cartesian grid spacing is set to 0.4\,mm in the 20-kHz, and to 0.16\,mm in the 80-kHz PSM simulations, while $NGLL=7$ in all shown \textsc{specfem3d} simulations. 
Normalized correlation $C$ between all traces in Fig.~\ref{fig:sphere_com}a are above 0.99 (Table~\ref{tab:corr_matrix}), while the correlation of the two traces in  \ref{fig:sphere_com}b is 0.997.

\begin{center}
    \centering
   \includegraphics[width=0.95\linewidth]{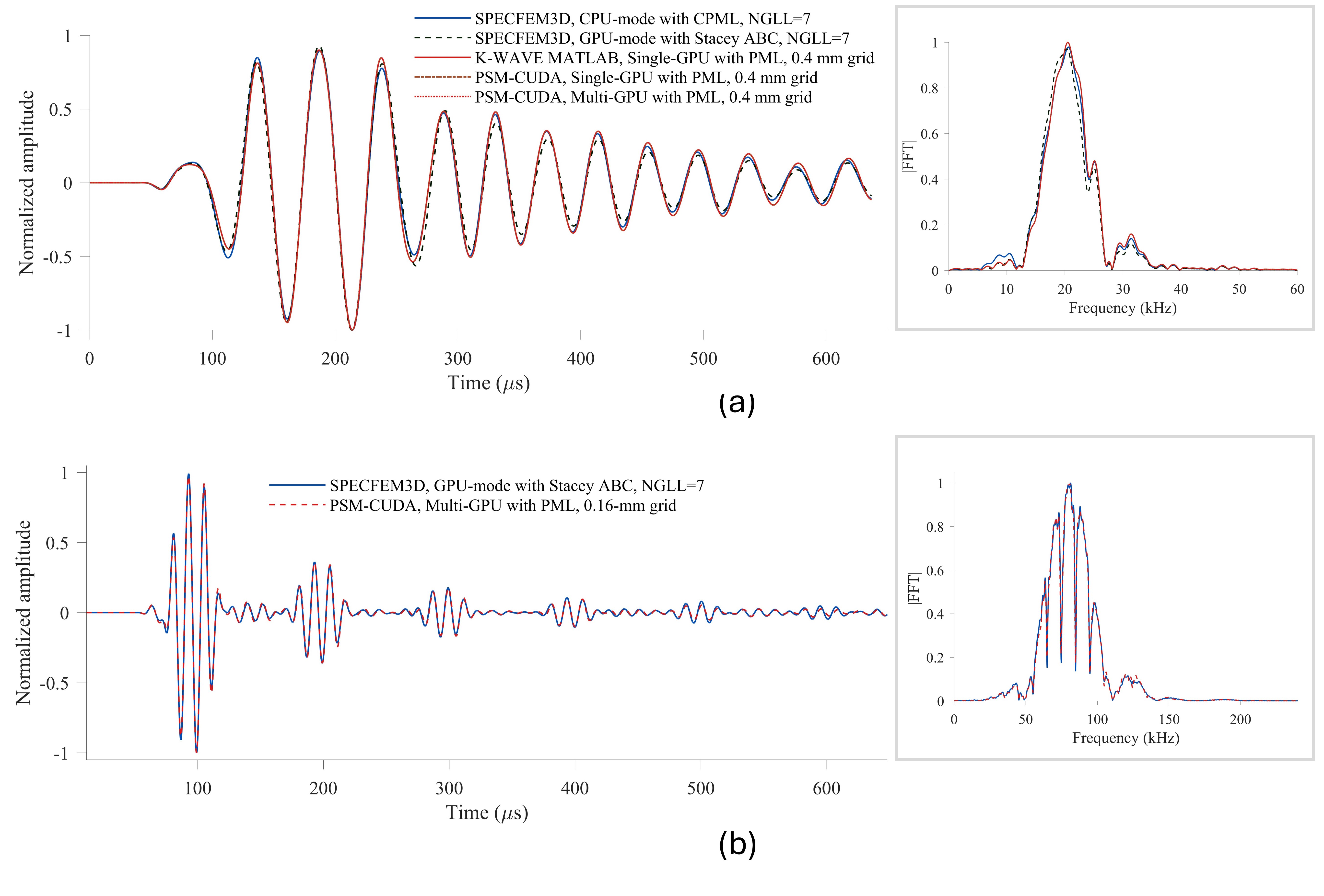}
\captionof{figure}{Pressure recorded at the sensor of the sphere model (as in Figure~\ref{fig:dolphin_workflow}b) for an incident plane wave: (a) 20\,kHz, comparing \textsc{specfem3d} with 7 GLL points per element under CPML (CPU) and Stacey (GPU) absorbing boundary conditions (ABC) against PSM in \textsc{Matlab} and \textsc{psm-cuda} in single- and multi-GPU configurations on a 0.4\,mm Cartesian grid; (b) 80\,kHz, same configurations with \textsc{specfem3d} (NGLL${}=7$) in GPU mode and multi-GPU \textsc{psm-cuda} on a 0.16\,mm voxel grid.}
\label{fig:sphere_com}
\end{center}

The highest was observed between the \textsc{matlab} and CUDA implementations of the PSM solver. This result is expected because both solvers implement the same numerical formulation, with the CUDA version differing primarily in domain decomposition, communication between GPUs and parallel execution, while preserving the underlying mathematical operations. Comparisons involving \textsc{specfem3d} exhibited slightly lower correlation coefficients, {presumably owing} to the different absorbing-boundary-condition formulations (CPML versus Stacey) and the inherent differences between the SEM finite-element and PSM voxel grid discretizations.

All simulations that follow result from either the multi-GPU CUDA implementation of \textsc{{\it k}-wave}, or from \textsc{specfem3d} with Stacey absorbing boundary conditions.  

\begin{table}[h]
\centering
  \fontfamily{ptm}\selectfont
  \renewcommand{\sfdefault}{ptm}
  \setlength{\tabcolsep}{4pt}
\caption{Normalized correlation of all possible pairs of traces from Fig.~\ref{fig:sphere_com}a.}
\label{tab:corr_matrix}
\begin{tabular}{lccccc}
\toprule
 & \makecell{SPECFEM3D\\PML} & \makecell{SPECFEM3D\\Stacey} & \makecell{K-WAVE\\MATLAB} & \makecell{PSM-CUDA\\Single-GPU} & \makecell{PSM-CUDA\\Multi-GPU} \\
\midrule
SPECFEM3D PML                  & 1.000 & 0.992 & 0.997 & 0.997 & 0.997 \\
SPECFEM3D Stacey               & 0.992 & 1.000 & 0.991 & 0.991 & 0.991 \\
K-WAVE MATLAB         & 0.997 & 0.991 & 1.000 & 1.000 & 1.000 \\
PSM-CUDA Single-GPU  & 0.997 & 0.991 & 1.000 & 1.000 & 1.000 \\
PSM-CUDA Multi-GPU   & 0.997 & 0.991 & 1.000 & 1.000 & 1.000 \\
\bottomrule
\end{tabular}
\end{table}

\subsection{Dolphin's head}\label{sec:dolphin_head_model} 

The dolphin-head model consists of a rectangular box measuring $42 \times 42 \times 66$ cm, containing the hexahedral mesh described in sec.~\ref{sec:carlosmesh}, or its 3D Cartesian grid counterpart. A plane wave is injected into all models exactly as described in sec.~\ref{sec:spheretest} (except $c_a=1480$ m/s, $\rho_a=1028$ kg/m$^3$), with the same time dependence as in sec.~\ref{sec:spheretest} but varying central frequency. Signals were recorded by two receivers, positioned as shown in Figure~\ref{fig:dolphin_workflow}, i.e., (i) within the water domain in front of the rostrum and (ii) in soft tissue at the approximate location of the right inner ear.

The cross validation of the PSM and SEM that we are about to describe consists of testing, first, the stability of both PSM and SEM results, independent from one another, with respect to changes in parameterization density, i.e. with growing $NGLL$ (SEM) or increasingly fine uniform Cartesian grid spacing (PSM). All SEM (and PSM) simulations in the following are based on the exact same mesh, the spatial resolution of SEM being therefore only controlled by the value of $NGLL$. 

As mentioned, experimental wave propagation data for such a complex biological medium do not exist and cannot realistically be obtained. In the absence of a ``ground-truth'' dataset to compare against, the discrepancy between independent numerical simulations is an indication of the numerical error that they might carry. Given the profound differences between PSM and SEM, it is at the very least unlikely that they should provide similar results in the presence of implementation errors or significant numerical noise. 



\subsubsection{Convergence/stability of PSM results with growing resolution}\label{sec:psmconv}

We conduct \textsc{{\it k}-wave} simulations on the dolphin-head model with uniform grid spacing from 1 to 0.4 mm, with decrements of 0.1 mm, and compare the results in figures \ref{fig:dolphinbeak}--\ref{fig:psmabcd}. Note that the same value of grid spacing translates to very different numbers of points-per-wavelength (PPW), depending on wave speed, which changes across the medium. If we take the lowest wave speed value found over the entire dolphin-head model, i.e. that of shear-wave velocity in soft tissues, grid density is still as high as 11~PPW at 0.9~mm and 25~PPW at 0.4~mm resolution, and only drops to 2.2 and 4~PPW at 100~kHz. PPW is much higher than that throughout the vast majority of the propagation medium, and for compressional waves vs. shear waves.

It is apparent from Fig.~\ref{fig:psmabcd} that the overall discrepancy between different-resolution simulation results decreases with increasing grid points, as expected, suggesting that, as resolution increases, numerical noise (which we expect to change in a random fashion across simulations) is reduced and modeled traces approximately converge towards one stable solution.

As a general rule, normalized correlation $C$ between all traces and the reference 0.4-mm ones are extremely high (never below $0.99784$ at the rostrum and  $0.88$ at the inner ear locations). Indeed, both Figs. \ref{fig:dolphinbeak} and \ref{fig:dolphinear} show that all traces are almost perfectly in phase; differences in amplitude are more visible, and are reflected by relatively large RMS differences: (up to 30\% when comparing the higher-frequency 0.9-mm traces to the 0.4-mm at the inner-ear, but always below 20 \% at resolutions of 0.7-mm and higher, both at rostrum and inner ear. (RMS difference is sensitive to both amplitude and phase, while normalized correlation is only affected by phase differences.)

Similarity between traces is generally higher within the water domain in front of the rostrum, where the modeled/recorded signal should be a straightforward combination of input sound and its direct reflection off the beak. At this location, the effects of complex anatomical structures are minor, and the ``coda'' of modeled traces accordingly less prominent.

At the at the approximate location of the inner ear, the recorded pressure signals also converge in both the time and frequency domains, although their response to grid refinement is more pronounced at higher frequencies. While the initial, relatively-large-amplitude segment of the signal is, again, very stable across grid resolutions, the 90- and 100-kHz traces at 1-mm and 0.9-mm are visibly different from the higher-resolution ones in both phase and amplitude (Fig.~\ref{fig:dolphinear}).

\begin{center}
    \centering
    \includegraphics[width=1\linewidth]{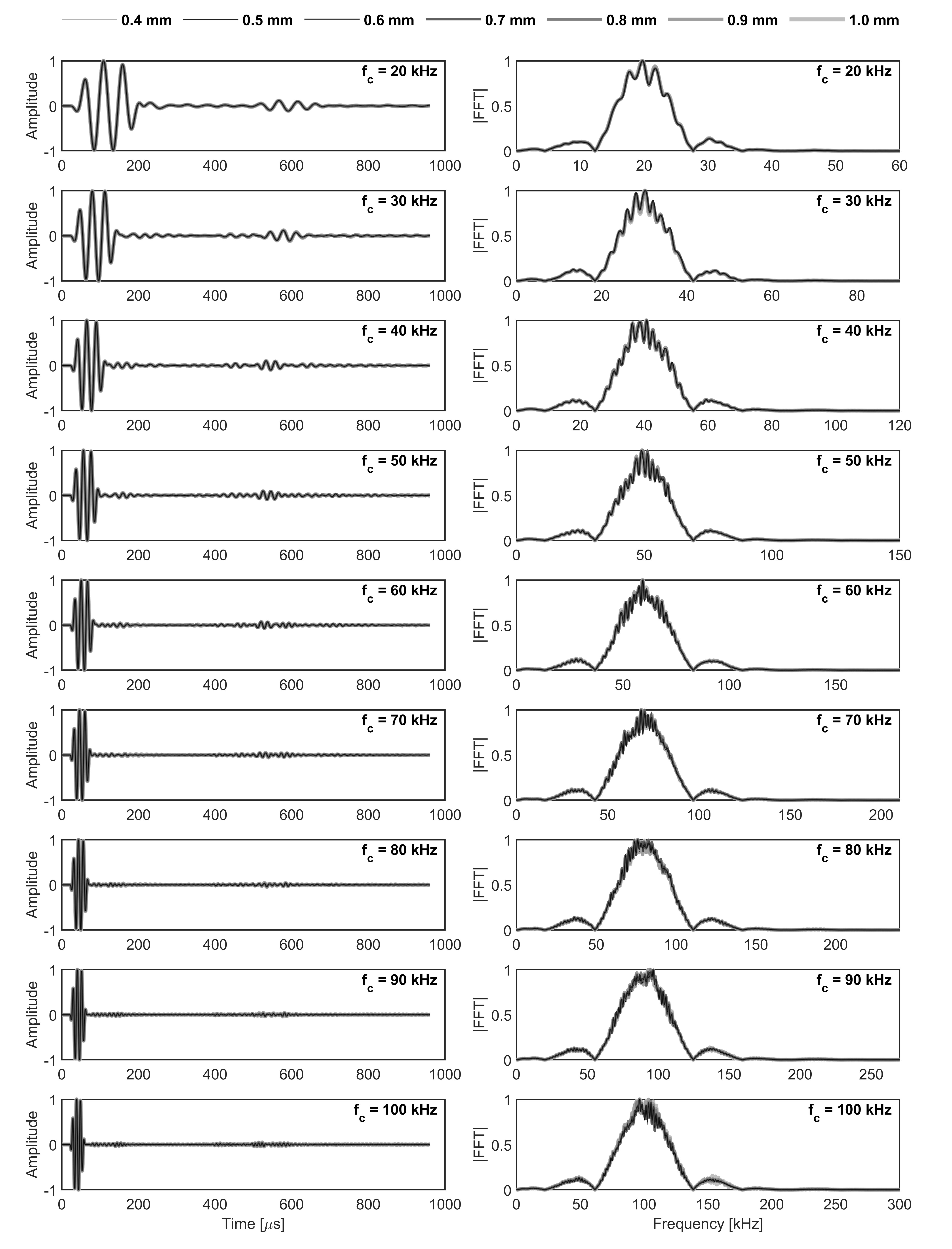}
   \captionof{figure}{Dolphin-head \textsc{psm-cuda} simulations with central frequencies ranging from (top to bottom) 20 kHz to 100 kHz, and grid spacing from 0.4 mm to 1 mm (different shades of grey). Waveforms recorded in front of the rostrum}
    \label{fig:dolphinbeak}
\end{center}

\begin{center}
    \centering
      \includegraphics[width=1\linewidth]{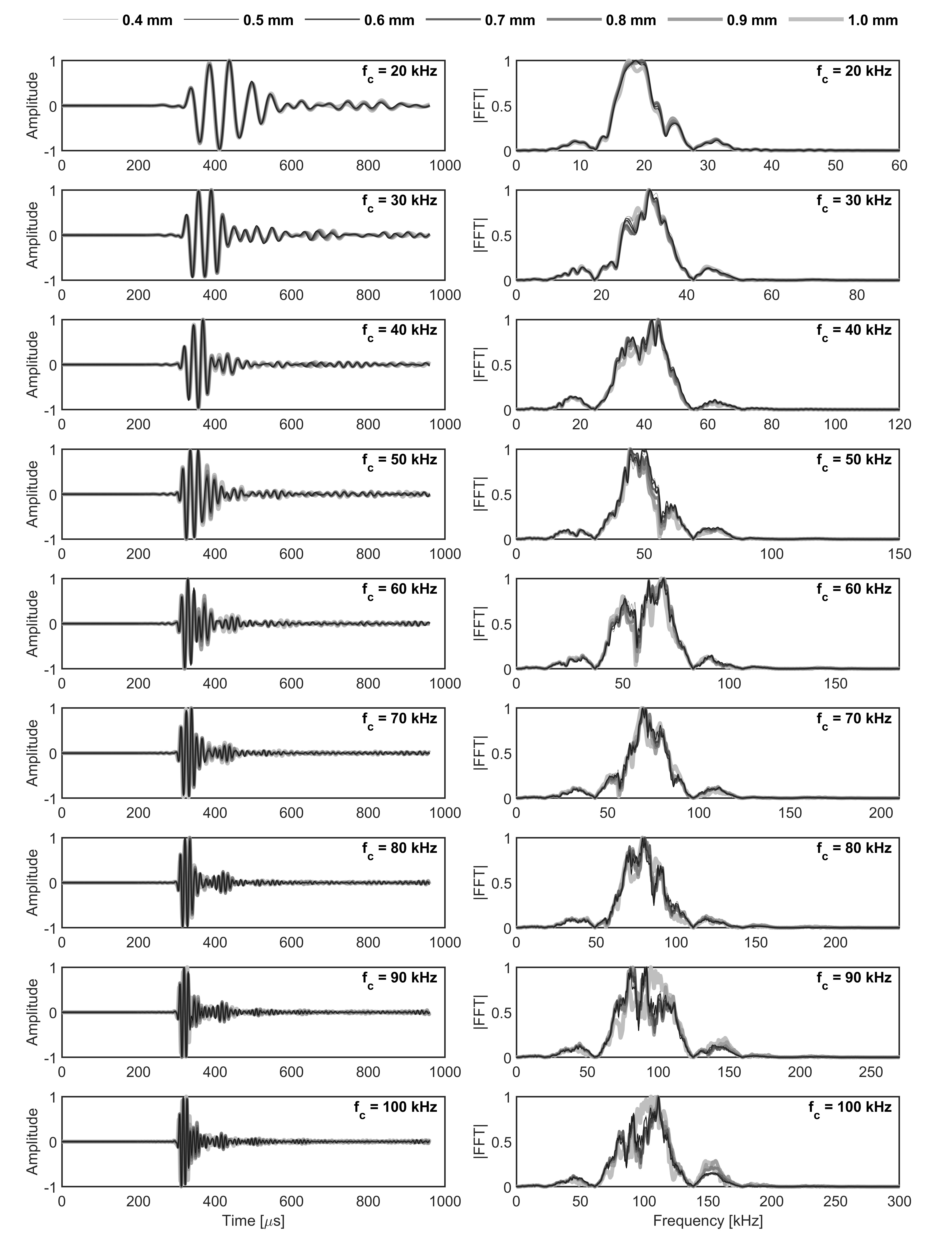}
    \captionof{figure}{Same as fig. \ref{fig:dolphinbeak}, but waveforms are recorded at the approximate location of the inner ear.}
    \label{fig:dolphinear}
\end{center}

\begin{center}
    \centering
     \includegraphics[width=0.9\linewidth]{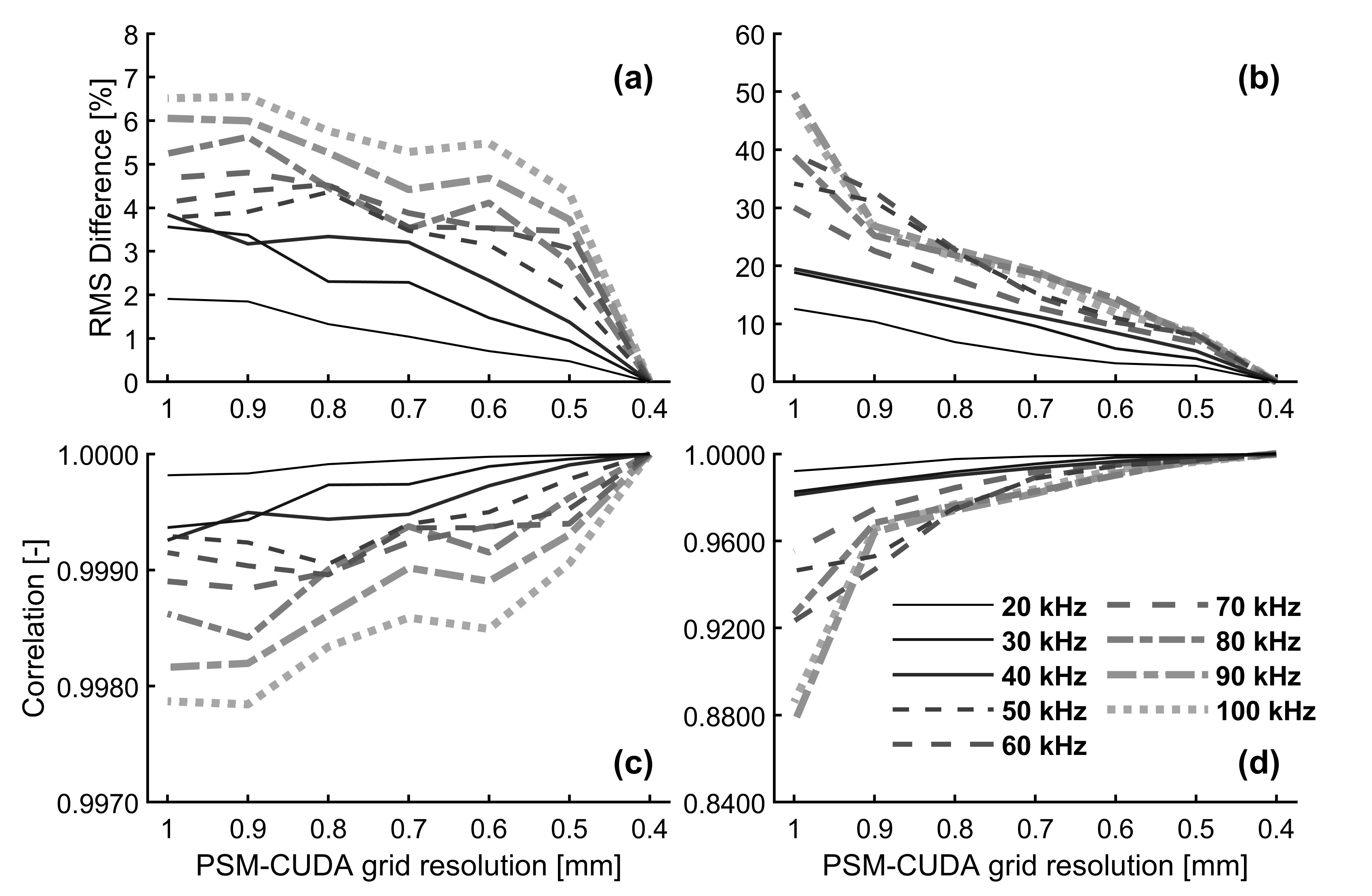}
    \captionof{figure}{Percent relative RMS difference (a,b) and normalized correlation (c,d) between the 0.4-mm resolution PSM dolphin-head simulation results, in front of the rostrum (a,c) and at the approximate inner-ear location (b,d), as functions of grid spacing.}
     \label{fig:psmabcd}
\end{center}

\subsubsection{Convergence/stability of SEM results with growing resolution}  

After conducting \textsc{specfem3d} simulations with $NGLL=$3, 4, 5 and 6, we compare the results in Figures \ref{fig:semconv1}--\ref{fig:semabcd}, same as we did in sec.~\ref{sec:psmconv} for PSM results. 

Our findings essentially mirror those of sec.~\ref{sec:psmconv}: again, discrepancy decreases with increasing $NGLL$, confirming the essential stability of high-resolution SEM results; again, traces recorded in front of the rostrum are overall very similar, while the coda of $NGLL=3$ and, to some extent, $NGLL=4$ high-frequency traces recorded at the inner ear differ significantly from the $NGLL=6$ reference (see the bottom of Fig.~\ref{fig:semconv2}). 
But, again, even this difference disappears if one compares the higher-resolution $NGLL=5$ and $NGLL=6$ results.

\subsubsection{PSM-SEM Comparison} 
PSM and SEM agree remarkably well with each other over the full frequency range modelled here (Figures \ref{fig:dolphi}, \ref{fig:dolph} and \ref{fig:comb}).  Once again, as expected, model similarity clearly grows with increasing resolution, confirming that simulation results tend to become stable once medium parameterization is sufficiently detailed. 

We explain the small mismatch in Figure~\ref{fig:comb}, at the low-frequency end of the 
20 kHz simulation, in terms of the plane-wave source implementation, since low-frequency wavelengths produce a small diffraction tail that disappears at higher frequencies (an issue identified during source implementation in \textsc{psm-cuda}, but not present in \textsc{specfem3d}). We also observe that the overall correlation values are lower at the inner ear recordings however still higher than 0.99 (in Figure~\ref{fig:dolph}). We are fairly confident that this small disagreement arises from geometric approximation error in the voxel grid at curved interfaces, as well as from the Stacey boundary conditions, which may produce small reflections at grazing angles. The frequency-domain results exhibit good agreement at both recording locations. Similar spectral minima

\begin{center}
    \centering
    \includegraphics[width=1\linewidth]{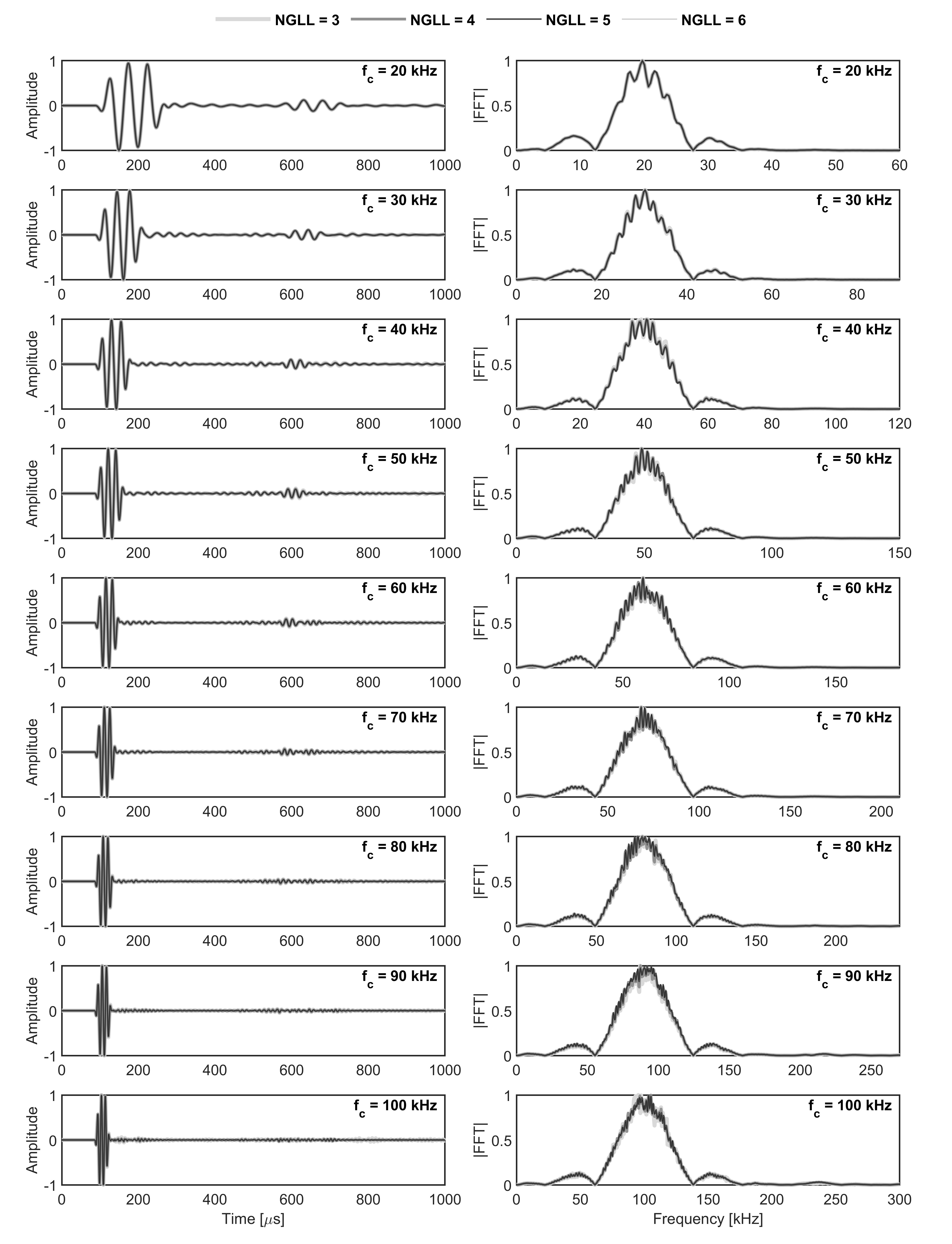}
    \captionof{figure}{Dolphin-head  \textsc{specfem3d} simulations with central frequencies ranging from (top to bottom) 20 kHz to 100 kHz, and $NGLL=4, 5, 6$ (different shades of grey). Waveforms recorded in front of the rostrum.}
    \label{fig:semconv1}
\end{center}

\begin{center}
    \centering
      \includegraphics[width=1\linewidth]{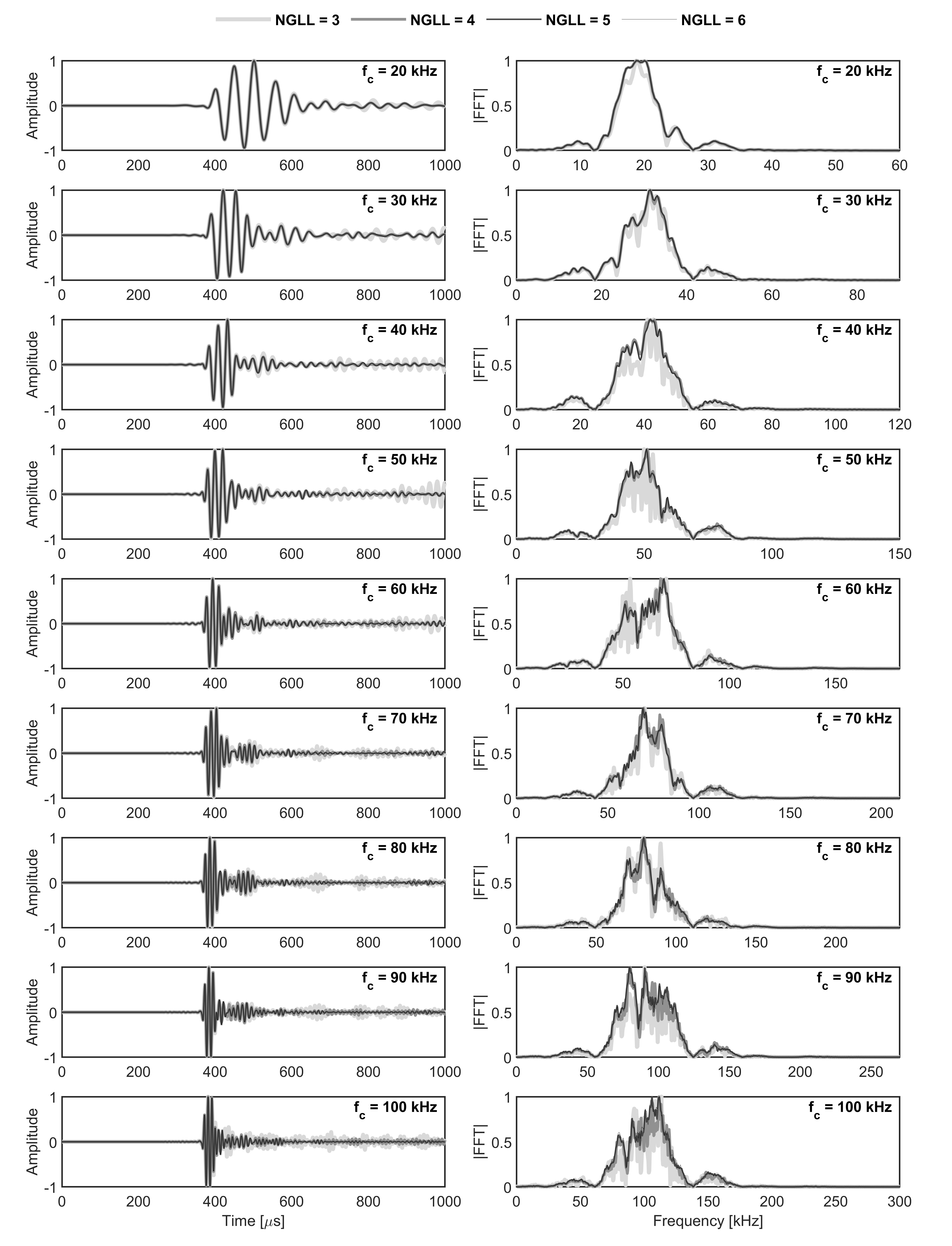}
   \captionof{figure}{Same as fig. \ref{fig:semconv1}, but waveforms are recorded at the approximate location of the inner ear.}
     \label{fig:semconv2}
\end{center}

\begin{center}
    \centering
     \includegraphics[width=0.9\linewidth]{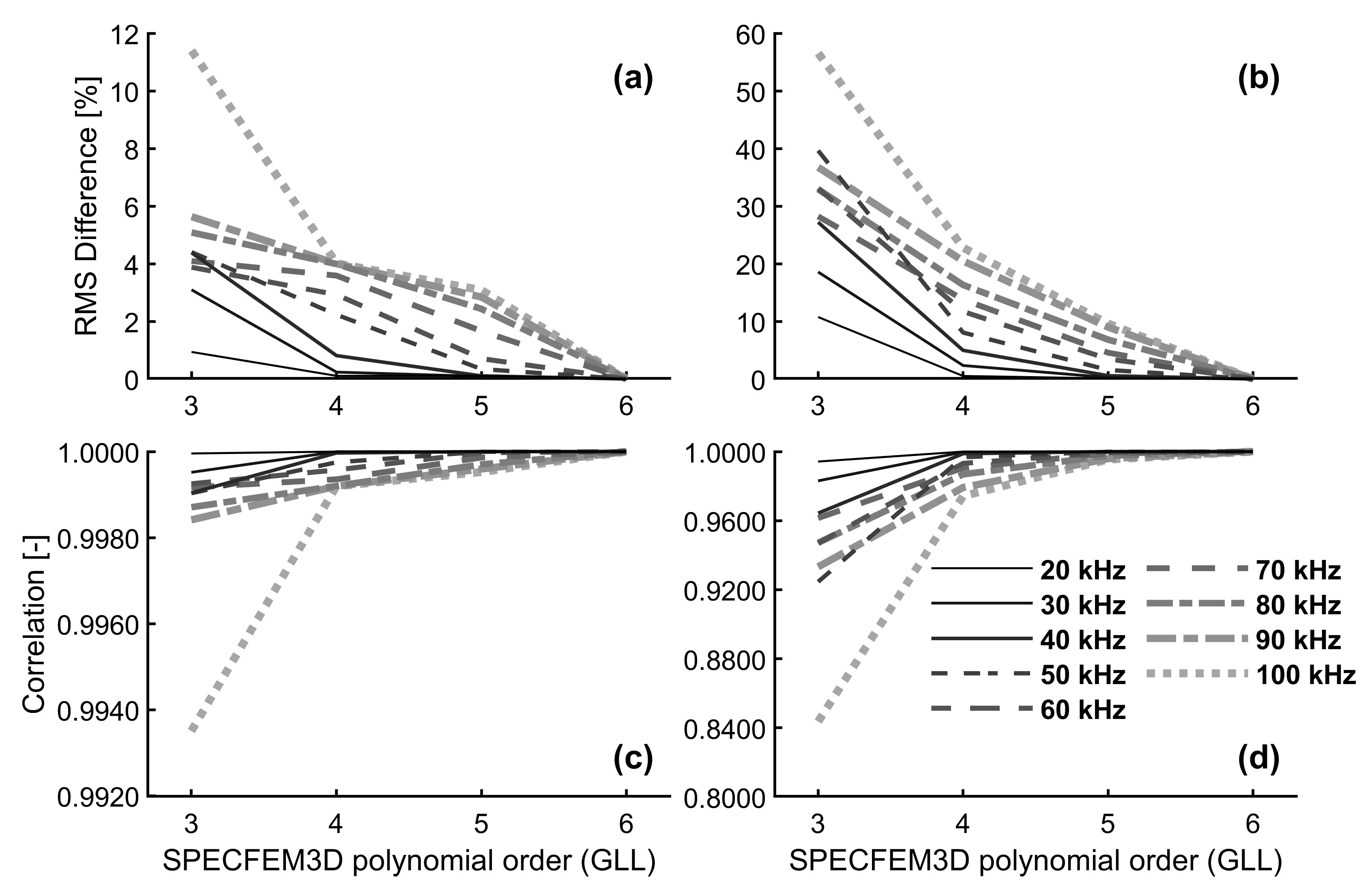}
    \captionof{figure}{Percent relative RMS difference (a,b) and normalized correlation (c,d) between the $NGLL=6$ SEM dolphin-head simulation results, in front of the rostrum (a,c) and at the approximate inner-ear location (b,d), as functions of grid spacing.}
     \label{fig:semabcd}
\end{center}

\noindent occur at consistent frequencies, which implies that one should be able to reliably identify the spectral ``notches''\citep[e.g.][ch. 7]{vanopstal16}, if any, of dolphin audition.

\subsection{Performance of PSM on various computational architectures}

Simulating ultrasound propagation through realistic biological models requires significant computational resources on cutting-edge high performance computing infrastructure. For instance, our \textsc{specfem3d}, $NGLL=6$ simulation of 100 kHz wave propagation (Figs. \ref{fig:semconv1} and \ref{fig:semconv2}) took 12 hours and 49 minutes of our real-world time, running in parallel on four NVIDIA H100 GPUs. Our highest resolution PSM simulation (0.4-mm grid spacing, Figs. \ref{fig:dolphinbeak} and \ref{fig:dolphinear}) took 4 hours and 36 minutes on five NVIDIA H100 GPUs.

To assess the performance of the \textsc{psm-cuda} version of \textsc{{\textit k}-wave} on one vs. two GPUs, we ran the same simulation on the simpler model of sec.~\ref{sec:spheretest} and on three different NVIDIA architectures, measuring performance as documented by Table~\ref{tab:kwave_gpu_bench}.

The NVIDIA L40S (Ada architecture \citep{nvidia2023l40s}) GPU is extraordinarily performant at single precision computation, but lacks NVLink\citep{NVLINK}, 
so that multi-GPU communication happens via PCIe (Peripheral Component Interconnect Express)\citep{lutz2020pump}. The A100 GPU\citep{turisini2023leonardo} is relatively low-speed in terms of the number of single-precision (FP32) TFLOPS (Tera floating-point operations per second) that it can carry out, but its memory bandwidth is higher than that of the L40S. Finally,  the latest Hopper architecture \citep{nvidia2022hopper,choquette2023nvidia} has medium (FP32)-TFLOPS performance but is well suited in terms of NVLink and memory bandwidth.

In practice, by monitoring the computational costs of our multi-GPU simulations we have found that NVLink and memory bandwidth are more important parameters than the number of TFLOPS, resulting in the L40S setup being outperformed by both the A100 and H100 ones, with H100 ultimately proving fastest. Albeit not perfect, the acceleration achieved by doubling the number of H100 GPUs (from one to 

\begin{center}
    \includegraphics[width=0.95\linewidth]{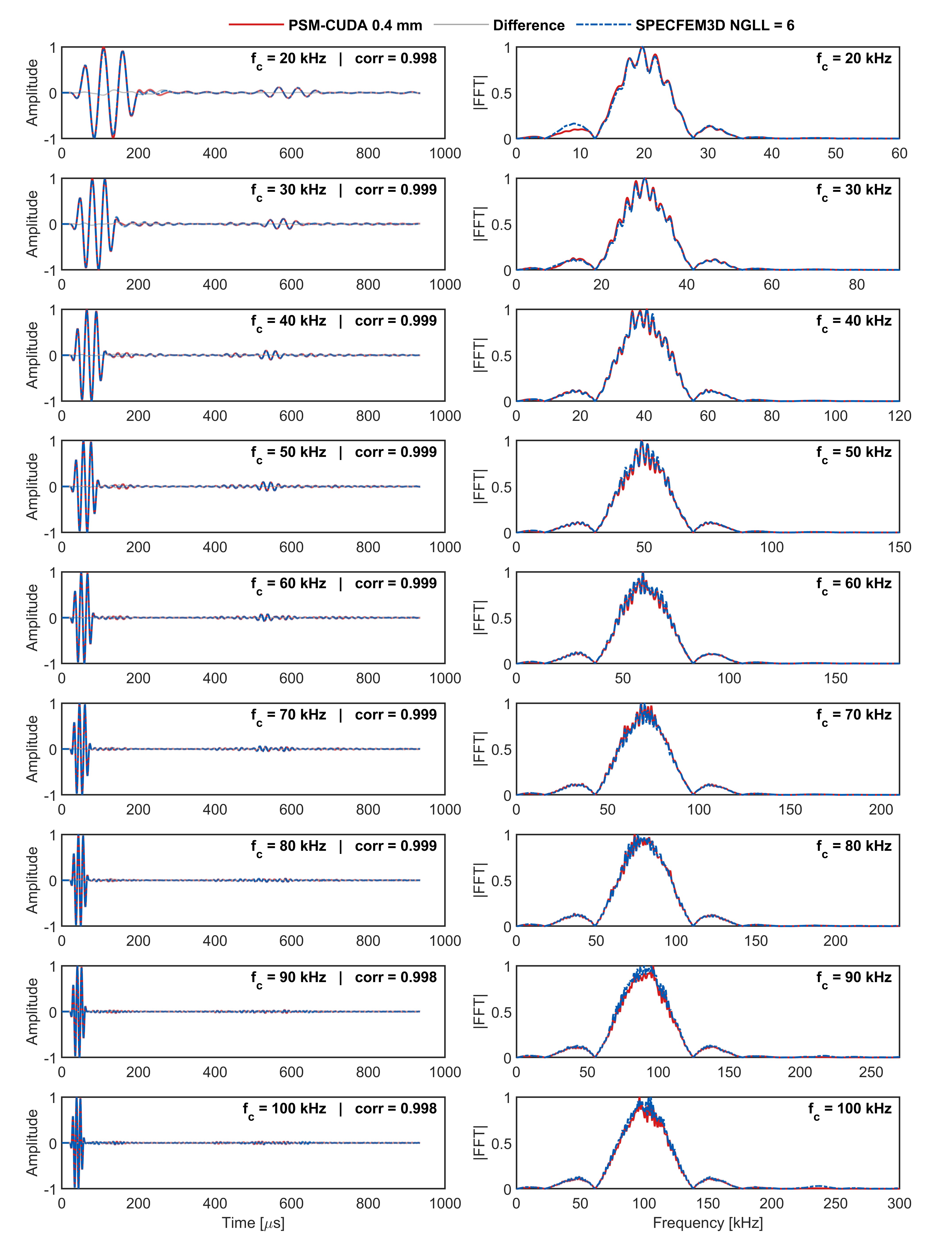}
   \captionof{figure}{Comparison of \textsc{specfem3d} ($NGLL = 6$)  and PSM-CUDA (0.4 mm grid spacing) waveforms recorded in front of the rostrum, with various central frequencies, as indicated.}
    \label{fig:dolphi}
\end{center}

\begin{center}
    \centering
      \includegraphics[width=0.95\linewidth]{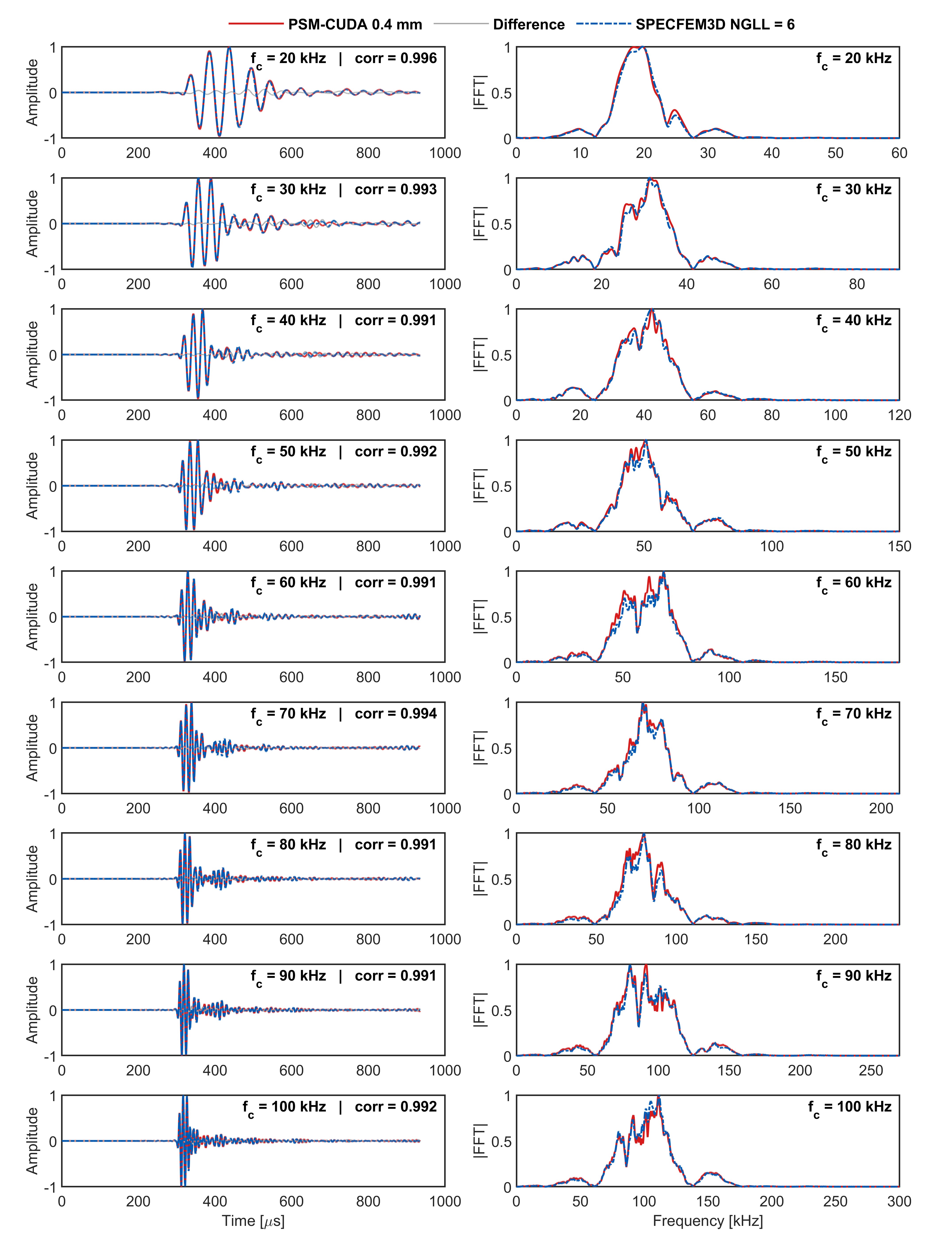}
    \captionof{figure}{Same as figure \ref{fig:dolphi} , waveforms recorded at the approximate inner-ear location.}
    \label{fig:dolph}
\end{center}

\begin{center}
    \includegraphics[width=0.82\linewidth]{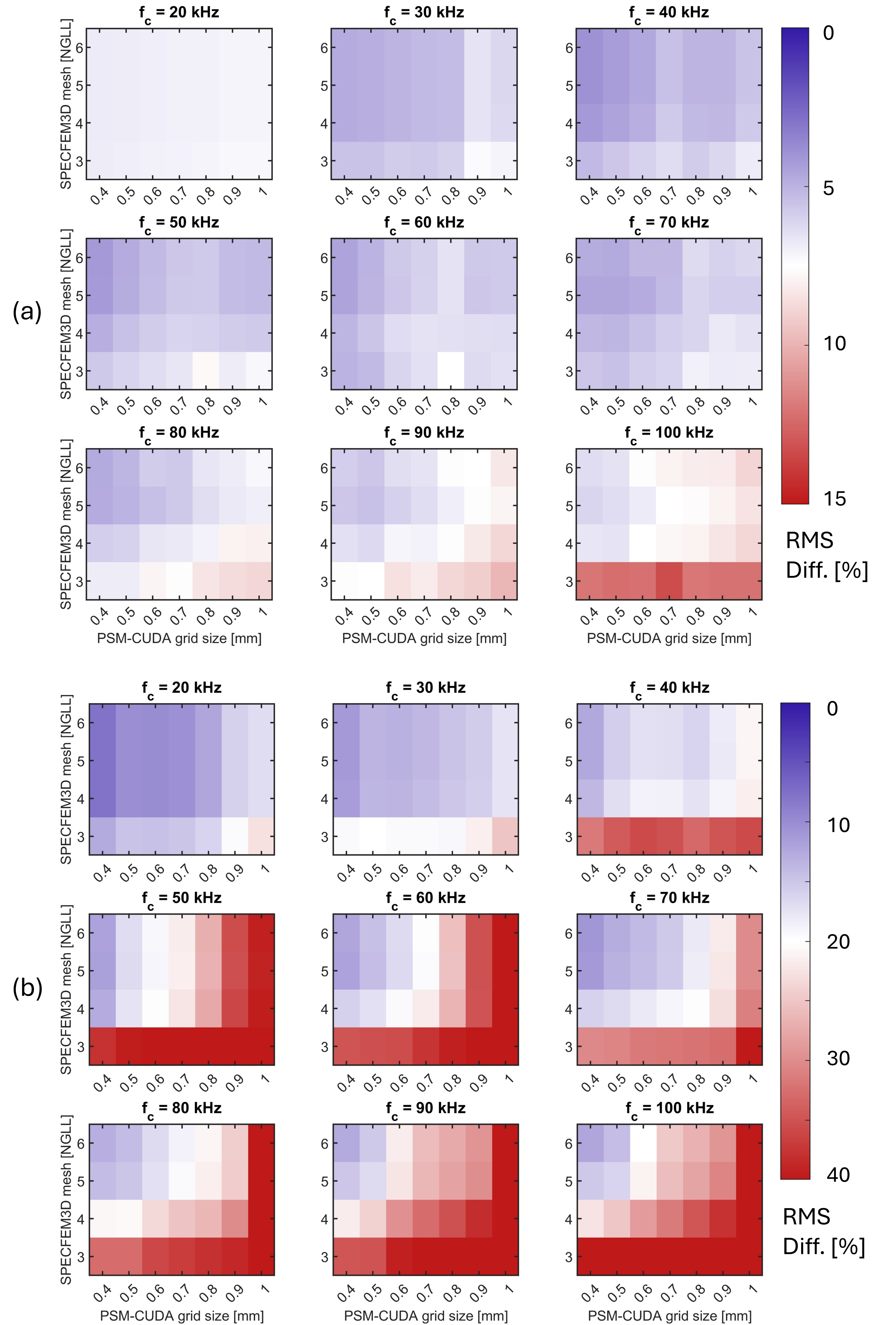}
   \captionof{figure}{Dolphin head model comparison study: RMS difference (in percent) for all cases, comparing \textsc{specfem3d} (NGLL varied from 3 to 6) against \textsc{psm-cuda} (voxel grid ranging from 1 mm to 0.4 mm), across central frequencies from 20 kHz to 100 kHz, recording the waveform at (a) in front of the rostrum and (b) inner-ear location.}
    \label{fig:comb}
\end{center}

\begin{table}[!htbp]
  \centering
  \fontfamily{ptm}\selectfont
  \renewcommand{\sfdefault}{ptm}
  \caption{\textsc{psm-cuda} per-time-step compute cost on three NVIDIA GPU
           architectures, measured in single- and dual-GPU configurations.}
  \label{tab:kwave_gpu_bench}
  \setlength{\tabcolsep}{4pt}
  \renewcommand{\arraystretch}{1.15}
  \begin{tabular}{lccc}
  \toprule
& \textbf{Ada Lovelace L40S} & \textbf{Ampere A100} & \textbf{Hopper H100} \\
& \makecell{(48\,GB GDDR6)}
& \makecell{(64\,GB HBM2e\\Leonardo CINECA)}
& \makecell{(80\,GB HBM3\\DGX-SXM5)} \\
\midrule
    Peak FP32 [TFLOPS]                & 91.6   & 22.4    &  67     \\
    Memory bandwidth [GB/s]           & 864    & 1{,}640 & 3{,}350 \\
    NVLink / interconnect [GB/s]      & --     & 200     & 900     \\
    \midrule
    Single GPU  Time-step [ms]           & 84.31  & 56.18   & 27.14   \\
    Dual GPU   Time-step [ms]           & 62.58  & 37.08   & 15.59   \\
    \midrule
    Speed-up, dual vs.\ single [ratio]       & 1.35$\times$  & 1.52$\times$  & 1.74$\times$ \\
    \bottomrule
  \end{tabular}
\end{table}

\begin{table}[!htbp]
  \centering
  \fontfamily{ptm}\selectfont
  \renewcommand{\sfdefault}{ptm}
  \caption{Computational cost of the \textsc{specfem3d} coupled acoustic--elastic
           solver as a function of the spectral-element polynomial order
           NGLL, evaluated on the same hexahedral dolphin-head mesh of
           12{,}676{,}714 elements (5{,}547{,}853 acoustic +
           7{,}128{,}861 elastic). All runs uses single precision on NVIDIA H100 80~GB HBM3}
  \label{tab:specfem_ngll}
  \footnotesize
  \setlength{\tabcolsep}{6pt}
  \renewcommand{\arraystretch}{1.150}
  \begin{tabular}{crrrrrrrrrrr}
    \toprule
    NGLL
      & \multicolumn{1}{c}{\makecell[c]{GLL mesh\\points}}
      & \multicolumn{1}{c}{\makecell[c]{$\Delta_t$}}
      & \multicolumn{1}{c}{\makecell[c]{$\Delta t_{\textrm {max}}$}}
      & \multicolumn{1}{c}{\makecell[c]{$N_t$}}
      & \multicolumn{1}{c}{\makecell[c]{$N_\mathrm{GPU}$}}
      & \multicolumn{1}{c}{\makecell[c]{VRAM\\used}}
      & \multicolumn{1}{c}{\makecell[c]{VRAM\\total}}
      & \multicolumn{1}{c}{\makecell[c]{$\tau$}}
      & \multicolumn{1}{c}{\makecell[c]{Wall}}
      & \multicolumn{1}{c}{\makecell[c]{GPU-h}} \\
      & \multicolumn{1}{c}{$\times 10^9$}
      & [ns]
      & [ns]
      &
      &
      & \multicolumn{1}{c}{[GB/GPU]}
      & \multicolumn{1}{c}{[GB]}
      & \multicolumn{1}{c}{[ms/step]}
      & \multicolumn{1}{c}{[h]}
      & \\
    \midrule
    3 & 0.10 & 2.0 & 6.55 & 550{,}000  & 4 &  8.9 & 320 &  7.80 &  1.20 &   4.8 \\
    4 & 0.35 & 2.0 & 3.60 & 550{,}000  & 4 & 18.8 & 320 & 17.05 &  2.62 &  10.5 \\
    5 & 0.81 & 2.0 & 2.25 & 550{,}000  & 4 & 38.0 & 320 & 33.88 &  5.19 &  20.76 \\
    6 & 1.59 & 1.5 & 1.50 & 733{,}333  & 4 & 72.7 & 320 & 62.96 & 12.84 & 51.36 \\
    \bottomrule
  \end{tabular}

  \vspace{0.4em}
  \footnotesize
  GPU-h $= \tau_\mathrm{wall} \cdot N_\mathrm{GPU}$.  $\tau$ is the mean
  wall time per time step. $\Delta t_{\textrm {max}}$ is the maximum stable time step
  reported by \textsc{specfem3d}.  VRAM~used is the per-GPU peak working set;
  VRAM~total is the aggregate capacity across all GPUs (80~GB per H100).
  Mesh load imbalance (NSPEC$_{\max}$/NSPEC$_{\min}$) is 1.70 for all four
  runs, reflecting SCOTCH's compensation for the different per-element cost
  of acoustic vs.\ elastic elements.
\end{table}

\begin{table}[!htbp]
  \centering
  \fontfamily{ptm}\selectfont
  \renewcommand{\sfdefault}{ptm}
  \caption{Computational cost of the \textsc{psm-cuda} elastic solver across grid
           resolutions. All models are running on single precision (FP32), NVIDIA H100 80~GB HBM3 GPUs with
           NCCL-based Z-slab decomposition. The 0.4~mm reference case
           required five GPUs to fit in memory; the remaining cases fit
           comfortably in four.}
  \label{tab:pstd_grid_resolution}
  \footnotesize
  \setlength{\tabcolsep}{4pt}
  \renewcommand{\arraystretch}{1.10}
  \begin{tabular}{lrrrrrrrrrr}
    \toprule
    Case
      & \multicolumn{1}{c}{Grid}
      & \multicolumn{1}{c}{Voxels}
      & $\Delta t$
      & \multicolumn{1}{c}{$N_t$}
      & \multicolumn{1}{c}{$N_\mathrm{GPU}$}
      & \multicolumn{1}{c}{\makecell[c]{VRAM\\used}}
      & \multicolumn{1}{c}{\makecell[c]{VRAM\\total}}
      & \multicolumn{1}{c}{$\tau$}
      & \multicolumn{1}{c}{Wall}
      & \multicolumn{1}{c}{GPU-h} \\
      &
      & \multicolumn{1}{c}{$\times 10^9$}
      & [ns]
      &
      &
      & \multicolumn{1}{c}{[GB/GPU]}
      & \multicolumn{1}{c}{[GB]}
      & \multicolumn{1}{c}{[ms/step]}
      & \multicolumn{1}{c}{[h]}
      & \\
    \midrule
    0.4~mm       & 1447$\times$1090$\times$1090 & 1.72  &  44.0 & 22{,}727 & 5 & 56.4 & 400 & 729.0 & 4.60 & 23.0 \\
    0.5~mm       & 1158$\times$872$\times$872   & 0.88  &  55.0 & 18{,}182 & 4 & 36.1 & 320 & 435.8 & 2.20 &  8.8 \\
    0.6~mm       &  965$\times$726$\times$726   & 0.51  &  66.0 & 15{,}152 & 4 & 20.9 & 320 & 193.0 & 0.81 &  3.3 \\
    0.7~mm       &  828$\times$622$\times$622   & 0.32  &  77.0 & 12{,}987 & 4 & 13.2 & 320 & 128.3 & 0.46 &  1.9 \\
    0.8~mm       &  724$\times$545$\times$545   & 0.215 &  88.0 & 11{,}364 & 4 &  8.9 & 320 & 119.7 & 0.38 &  1.5 \\
    0.9~mm       &  643$\times$485$\times$485   & 0.151 & 100.0 & 10{,}000 & 4 &  6.2 & 320 &  82.3 & 0.23 &  0.9 \\
    1.0~mm       &  578$\times$436$\times$436   & 0.110 & 100.0 & 10{,}000 & 4 &  4.5 & 320 &  47.3 & 0.13 &  0.5 \\
    \bottomrule
  \end{tabular}

  \vspace{0.4em}
  \footnotesize
  GPU-h $= \tau_\mathrm{wall} \cdot N_\mathrm{GPU}$ in GPU-hours.  $\tau$ is
  the mean wall time per time step.  VRAM~used is the per-GPU peak working
  set reported by the solver; VRAM~total is the aggregate capacity across
  all GPUs in the run (80~GB per H100). 
\end{table}


\noindent two) is quite good, with a single time step being completed 1.74 times more rapidly. This performance losses is presumably accounted for by the GPU communication via NCCL.

We next use the DGX-SXM5 (\citep{DGX}) H100 node available to us as reference, to analyze the performance of both SEM and PSM at more demanding dolphin-head simulations. Our findings are summarized in Tables~\ref{tab:specfem_ngll} and~\ref{tab:pstd_grid_resolution}. Most importantly, a Cartesian grid of $1.72 \times 10^9$ nodes (i.e. the dolphin-head model, with 0.4-mm resolution)
costs 729~ms per timestep on five GPUs. In comparison, a hexahedral mesh of the dolphin-head model with $NGLL=6$, equivalent to $1.59 \times 10^9$ grid points, takes only 62.96~ms per time step. 

For a more practical evaluation of both methods' performance, however, we also need to consider the actual simulated time-increment in seconds, corresponding to one time step in both simulations. In the SEM case, based on the CFL condition, we conservatively set time step $\Delta t =\frac{\Delta t_{\textrm {max}} }{\zeta} $ to
\begin{equation}
\Delta t_{\textrm {max}} = \frac{\Delta x}{V_{\textrm {max}}},
\end{equation}
where $V_{\textrm {max}}$ is the highest value of wave speed ($V_P$) throughout the entire model, and $\Delta x$ is the smallest spacing between neighbouring GLL points (which depends of course on the value of $NGLL$ for a given simulation), found at the smallest hexahedron edge of the entire mesh ($h_{\textrm{min}}$). The value of $\zeta$ is varied from 3.25 to 1.

Our $NGLL=6$ SEM simulation accordingly takes $6.67 \times 10^5$ time steps to model 1 ms of wave propagation, while our \textsc{psm-cuda} version of \textsc{{\it k}-wave} achieves that after only $2.2 \times 10^4$ time steps. Comparing our final GPU budget, we see that \textsc{psm-cuda} consumed approximately 23 GPU-hours, whereas \textsc{specfem3d} needed approximately 46.71 GPU-h.

This informal comparison does not allow us to conclude whether either approach is ``more performant'' than the other, nor was that our objective. Importantly, however, our results suggest that \textsc{specfem3d} and the \textsc{psm-cuda} version of \textsc{{\it k}-wave} are comparable in terms of both accuracy and speed.

\section{Conclusions} 
\noindent Recent literature \citep{igel2017computational,hejazi2020contribution,ali2025pseudo,garcia2026feasibility} suggests that the pseudospectral and spectral element methods (PSM and SEM, respectively), together with the finite element method \citep{wei2024validated} are effective ways to model elastic wave propagation (i.e. ultrasound) through the complex anatomy a dolphin. Recent work by our team has emphasized the potential contribution of wave propagation modeling to the problem of understanding the mysterious sound (echo)localization system of cetaceans; but being able to model elastic waves through biological tissues is relevant to multiple aspects of the life sciences/medical physics \citep{bachmann2020source,marty2024transcranial,treeby2012modeling,guasch2020full}.

Because the vestigial ear canals of dolphins are occluded, and their inner ears inaccessible, numerical modeling is practically the only way to ``measure'' sound as received by their cochlea. For related reasons, modeled sound cannot be easily compared to experimental data (laboratory experiment are hard to conduct on quickly decomposing post-mortem specimens, while experiments on live dolphins require dedicated infrastructure and time-consuming training). To convincingly validate our models, then, we computed numerical wave propagation with two independent, completely different approaches -- namely, the PSM and SEM as implemented in earlier work by our team [\citep{ali2025pseudo} and \citep{garcia2026feasibility}, respectively] -- and checked the similarity of the results. The nature of numerical error emerging in both solvers is quite different (it propagates at the same speed as modeled waves in SEM; leaks through the entire model at each PSM step), so that the similarity between modeled signals across methods is a legitimate measure of model accuracy.

Once a sufficiently high level of resolution is reached, we find indeed a strikingly high correlation, confirming that both approaches are reliable enough to provide a ``ground-truth'' database of dolphin head-related impulse responses. This result is somewhat surprising, given: (i) reported difficulties in implementing fully absorbing boundary conditions in SEM; (ii) the relatively low minimum values of scaled Jacobian (Table~\ref{tab:mesh_stats}) in our hexahedral mesh\citep{garcia2026feasibility}; (iii) the difficulty of modeling fluid-solid interfaces \citep[e.g.,][]{van2002finite} in PSM, where no continuity conditions are prescribed at interface within the simulation domain. Our findings imply that, in practice, both PSM and SEM successfully deal with all these challenges.

We evaluated the computational performance of both \textsc{specfem3d} and of our own CUDA version of \textsc{{\it k}-wave}, which effectively runs on multiple GPUs. The cost of one simulation, at the maximum level of resolution considered here, is roughly 51 GPU-hours with \textsc{specfem3d} and 23 GPU-hours with \textsc{{\it k}-wave}. At the frequency range we considered, however, our analysis shows that models become quite stable with changing resolution, beyond 0.6 mm in PSM and 5 Gauss-Lobatto-Legendre points in \textsc{specfem3d}, which reduces the costs to about 21 and 9 GPU-hours, respectively. While computational costs are comparable, the PSM approach has the great advantage of bypassing the hexahedral-mesh bottleneck.

The level of resolution that we have been able to achieve in this study is superior to that of earlier simulations of elastic wave propagation through cetacean anatomy. Still, our access to high-performance computational infrastructure was not sufficient to model even more complex features, such as the fine 3D structures of the tympano-periotic complex, or the extremely sharp discontinuities in elastic parameters between biological tissues and air, associated with the sinuses of live animals. Such features have been ``smoothed out'' in the hexahedral mesh we used \citep{garcia2026feasibility}, but will be taken into account in our future work. Should \textsc{psm-cuda} prove powerful enough to model very fine structure and high contrasts with the same accuracy as \textsc{specfem3d} and no explosion of the necessary computational budget, that would be our go-to choice for future practical applications, where many simulations will have to be conducted on multiple anatomy models.

\bibliographystyle{elsarticle-num}

\bibliography{myrefs}

\section*{Acknowledgments}
\noindent We are particularly grateful to the developers of both \textsc{specfem3d} and \textsc{{\it k}-wave}, for their efforts to share their softwares with the scientific community.

This work has benefitted from our conversations with Andrea Colombi, Heiner Igel and Jeroen Tromp, and their insightful advice on numerical wave propagation solvers and benchmarks. We are also grateful to Dr. Nicola Praticelli for configuring the DGX node with NVIDIA H100 and L40 GPUs, made available by the University of Padua through the UPSCALE supercomputing facility. We additionally made use of the CINECA supercomputing center: the present work would not have been possible without access to their Leonardo Booster partition under ISCRA-C (project id: IsCd4\_Biosonar). We thank Sandro Mazzariol, Steffen De Vreese, Jean-Marie Graïc, Ksenia Orekhova, and the DIAPHONIA (project PCI2022-135022-2 funded by the JPI Ocean program) network for constant feedback on the anatomy and hearing of dolphins and for preparing the stereolithography (STL) files.

Our research is funded by the EU Horizon Europe research and innovation programme under the Marie Sklodowska-Curie Grant Agreement No. 101119769 (SEASOUNDS Doctoral Network), and from the National Recovery and Resilience Plan (NRRP), Mission 4, Component 2, Investment 1.1, Call for tender No. 1409 published on 14.9.2022 by the Italian Ministry of University and Research (MUR), funded by the European Union - NextGenerationEU - project title: SWIM: aSsessing the Impact of offshore Wind turbines on Marine mammals in the Adriatic sea - CUP C53D23010230001- Grant Assignment Decree No. 1388 adopted on 01.09.2023 by the Italian Ministry of University and Research (MUR).

\appendix
\section*{Appendix A: time-stepping equation for the pseudo-spectral method}
\label[appendix]{app:proof_section} 
\noindent To prove eq. (\ref{tabei5}), start with the result\citep[section 3.3 of][]{mickens94,mickens15} 
\begin{enumerate}
\item Let $f_1(j), f_2(j), \hdots, f_n(j)$ be a set of $n$ functions of integer argument $j$. If 
$f_1(i), f_2(i), \hdots, f_n(i)$ are {\it linearly dependent}, then their 
 so-called ``Casorati determinant'' equals zero for all $j$, i.e.\citep[section 3.2 of][]{mickens15},
\begin{equation}
\begin{vmatrix}
f_1(j) & f_2(j) & \hdots & f_n(j)\\
f_1(j+1) & f_2(j+1) & \hdots & f_n(j+1)\\
\vdots & \vdots & \ddots & \vdots \\
f_1(j+n) & f_2(j+n) & \hdots & f_n(j+n)
\end{vmatrix}
=0.
\end{equation}
\item Based on the previous result, if $n$ linearly independent functions $f_1(j), f_2(j)$, etc. are given, one can always construct a difference equation having all those functions as solutions. The equation, in the unknown function $y(j)$, 
is simply\citep[section 4.3 of][]{mickens15}.)
 
\begin{equation}
\begin{vmatrix}
y(j) & f_1(j) & f_2(j) & \hdots & f_n(j)\\
y(j+1) & f_1(j+1) & f_2(j+1) & \hdots & f_n(j+1)\\
\vdots & \vdots & \vdots & \ddots & \vdots \\
y(j+n) & f_1(j+n) & f_2(j+n) & \hdots & f_n(j+n)
\end{vmatrix}
=0.
\end{equation}
\end{enumerate}

Now, this result can be applied\citep{mickens94} to the harmonic-oscillator eq. (\ref{tabei3}): which is a second-order ordinary differential equation and, as such, has two linearly independent analytical solutions,
\begin{equation}
f_1(j)
= \mbox{e}^{i c |{\bf k}| t_j }
\end{equation}
and
\begin{equation}
f_2(j)
= \mbox{e}^{-i c |{\bf k}| t_j }.
\end{equation}
It follows that
\begin{equation}
\begin{vmatrix}
y(j) & \mbox{e}^{i c |{\bf k}| t_j } & \mbox{e}^{-i c |{\bf k}| t_j } \\
y(j+1) & \mbox{e}^{i c |{\bf k}| t_{j+1} } & \mbox{e}^{- i c |{\bf k}| t_{j+1} }\\
y(j+2) & \mbox{e}^{i c |{\bf k}| t_{j+2} } & \mbox{e}^{-i c |{\bf k}| t_{j+2} } 
\end{vmatrix}
=0.
\end{equation}
After some algebra, this collapses to
\begin{equation}\label{diffeq1}
y(j+2) 
-2 y(j+1) \cos(c |{\bf k}| \delta t)
+y(j)=0,
\end{equation}
where $\delta t=t_{j+1}-t_j$. 
Invoking the trigonometric identity
$\cos(x) = 1 - 2\sin^2\left(\frac{x}2\right)$, (\ref{diffeq1}) takes the form
\begin{equation}\label{diffeq1}
y(j+2) 
-2 y(j+1) \left[1 - 2 \sin^2\left(\frac{c |{\bf k}| \delta t}2\right) \right]
+y(j)=0,
\end{equation}
or, dividing everything by $ 4 \sin^2\left(\frac{c |{\bf k}| \delta t}2\right)$,
\begin{equation}\label{diffeq2}
\frac{y(j+2) -2 y(j+1) +y(j)}{4 \sin^2\left(\frac{c |{\bf k}| \delta t}2\right) }
+y(j+1)
=0,
\end{equation}
After replacing $y(j)$ with $p({\bf k},t_j)$, letting $t_{j+2}=t+\delta t$, $t_{j+1}=t$, etc., and multiplying everything by $c^2 |{\bf k}|^2$, we end up precisely with eq. (\ref{tabei5}).
\end{document}